\documentclass[a4paper,fleqn]{cas-dc}

\usepackage[numbers]{natbib}

\usepackage{tabularx}    % Add to preamble
\usepackage{adjustbox}   % Add to preamble
\usepackage{algpseudocode}
\usepackage{graphicx}
\usepackage{listings}
\usepackage{xcolor}  % optional, for colored code

\usepackage{hyperref}

\definecolor{codegray}{rgb}{0.5,0.5,0.5}
\definecolor{codegreen}{rgb}{0,0.6,0}
\definecolor{codepurple}{rgb}{0.58,0,0.82}
\definecolor{backcolour}{rgb}{0.97,0.97,0.97}

\lstdefinestyle{pythonstyle}{
    language=Python,
    basicstyle=\footnotesize\ttfamily,     % font size and type
    keywordstyle=\color{blue}\bfseries,    % keywords in blue
    commentstyle=\color{green!50!black},   % comments in green
    stringstyle=\color{codepurple},              % strings in red
    numberstyle=\tiny\color{gray},         % line numbers
    stepnumber=1,
    numbersep=5pt,
    showstringspaces=false,
    breaklines=true,                        % automatic line breaks
    breakatwhitespace=true,
    frame=single,                            % adds box
    framerule=0.5pt,
    backgroundcolor=\color{gray!10},        % light gray background
    captionpos=b,                            % caption at bottom
    xleftmargin=5pt,
    xrightmargin=5pt,
}

\usepackage{caption}  % for caption customization

\usepackage{tikz}
\usetikzlibrary{
  shapes.geometric,    % for rounded rectangles, cylinders
  arrows.meta,         % for -Latex arrowheads
  positioning,         % for relative positioning (below=of ...)
  calc,                % coordinate calculations
  shapes.symbols,      % if you want extra shape flexibility
  decorations.pathmorphing, % optional: for fancy arrows/lines
  fit,                 % optional: grouping boxes
  backgrounds           % optional: shading or layer control
}

\usepackage{booktabs}  % for \toprule, \midrule, \bottomrule
\usepackage{array}     % for column formatting
\usepackage{capt-of} % for footnotes in captions
\usepackage{algorithm}
\def\tsc#1{\csdef{#1}{\textsc{\lowercase{#1}}\xspace}}
\tsc{WGM}
\tsc{QE}
\tsc{EP}
\tsc{PMS}
\tsc{BEC}
\tsc{DE}
\begin{document}
\let\WriteBookmarks\relax
\def\floatpagepagefraction{1}
\def\textpagefraction{.001}
%\shorttitle{Research opportunities at LLM–control interface}
%\shorttitle{}
%\shortauthors{K. Nosrati et~al.}
%\shortauthors{}

\title[mode=title, size=\LARGE]{When control meets large language models: From words to dynamics}
\tnotemark[1]

\tnotetext[1]{This work was supported by the European Union's Horizon Europe research and innovation programme under the grant agreement No 101120657, project ENFIELD (European Lighthouse to Manifest Trustworthy and Green AI), by the Estonian Research Council (grant No. PRG1463), and by the Estonian Centre of Excellence in Energy Efficiency, ENER (grant TK230) funded by the Estonian Ministry of Education and Research.}

%\tnotetext[2]{The second title footnote which is a longer text matter to fill through the whole text width and overflow into another line in the footnotes area of the first page.}

\author[1]{Komeil Nosrati}
\cormark[1]
%\ead{komeil.nosrati@taltech.ee}

%\credit{Conceptualization of this study, Methodology, Software}

%\address[1]{, Street 129, 1043 NX Amsterdam, The Netherlands}
\affiliation[1]{organization={Department of Computer Systems, Tallinn University of Technology},
                postcode={12618}, 
                city={Tallinn},             
%               citysep={}, % Uncomment if no comma needed between city and postcode
         %       state={},
                country={Estonia}}

\author[1]{Aleksei Tepljakov}

\author[2]{Juri Belikov}
%\ead{juri.belikov@taltech.ee}
%\ead[URL]{https://www.university.org}

%\credit{Data curation, Writing - Original draft preparation}

\affiliation[2]{organization={Department of Software Science, Tallinn University of Technology},
                postcode={12618}, 
                city={Tallinn},             
%               citysep={}, % Uncomment if no comma needed between city and postcode
             %   state={},
                country={Estonia}}

\author[1]{Eduard Petlenkov}
%\cormark[2]
%\fnmark[1,3]
%\ead{eduard.petlenkov@taltech.ee}
%\ead[URL]{www.campus.in}

%\cortext[cor1]{}
%\cortext[cor2]{Corresponding author}
%\cortext[cor1]{Corresponding author (\includegraphics[height=1.5ex]{thumbnails/cas-email.jpeg}~komeil.nosrati@taltech.ee)}
\cortext[cor1]{Corresponding author: \texorpdfstring{\includegraphics[height=1.5ex]{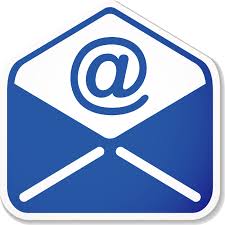}~komeil.nosrati@taltech.ee}{komeil.nosrati@taltech.ee}}
%\fntext[fn1]{This is the first author footnote, but is common to third author as well.}
%\fntext[fn2]{Another author footnote, this is a very long footnote and it should be a really long footnote. But this footnote is not yet sufficiently long enough to make two lines of footnote text.}

%\nonumnote{This note has no numbers. In this work we demonstrate $a_b$ the formation Y\_1 of a new type of polariton on the interface between a cuprous oxide slab and a polystyrene micro-sphere placed on the slab.}

\begin{abstract}
While large language models (LLMs) are transforming engineering and technology through enhanced control capabilities and decision support, they are simultaneously evolving into complex dynamical systems whose behavior must be regulated. This duality highlights a reciprocal connection in which prompts support control system design while control theory helps shape prompts to achieve specific goals efficiently. In this study, we frame this emerging interconnection of LLM and control as a bidirectional continuum, from prompt design to system dynamics. First, we investigate how LLMs can advance the field of control in two distinct capacities: directly, by assisting in the design and synthesis of controllers, and indirectly, by augmenting research workflows. Second, we examine how control concepts help LLMs steer their trajectories away from undesired meanings, improving reachability and alignment via input optimization, parameter editing, and activation-level interventions. Third, we look into deeper integrations by treating LLMs as dynamic systems within a state-space framework, where their internal representations are closely linked to external control loops. Finally, we identify key challenges and outline future research directions to understand LLM behavior and develop interpretable and controllable LLMs that are as trustworthy and robust as their electromechanical counterparts, thereby ensuring they continue to support and safeguard society.
\end{abstract}

%\includegraphics{Fig4_3DMeetr1.png}

%\item Traces the evolution of AI–control synergy from cybernetics to LLMs
%\item Introduces three paradigms for interactions between LLMs and control systems
%\item Shows how LLMs augment control design, optimization, and experimentation
%\item Explains how control theory improves LLM reliability and alignment
%\item Outlines challenges and future directions for controllable, trustworthy LLMs



\maketitle
\section{Introduction}
\begin{figure}
\centering
\includegraphics[width=\columnwidth]{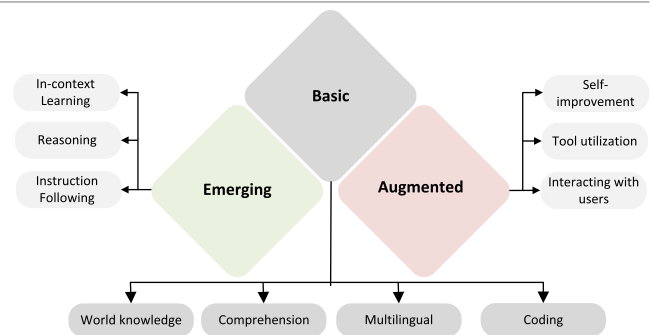}
\caption{Capabilities of LLMS.}
\label{fig_11}
\end{figure}

Rapid advances in large language models (LLMs) have reshaped artificial intelligence (AI)~\cite{Intro1}. Since the introduction of the Transformer architecture in 2017~\cite{Intro2}, LLMs have transformed natural language processing (NLP), progressing from basic text understanding and generation to sophisticated reasoning and intelligent interaction. Beyond basic skills in comprehension, world knowledge, coding, and multilingual understanding, LLMs exhibit emergent abilities absent in smaller pre-trained language models (PLMs)~\cite{Intro3}. As seen in Fig.~\ref{fig_11}, these include (i) in-context learning, where models acquire new tasks from a few examples at inference; (ii) instruction following, which enables models to perform novel tasks without explicit examples; and (iii) multi-step reasoning, in which models complete complex tasks through intermediate steps, as in chain-of-thought (CoT) prompting~\cite{Intro4}. LLMs can also be augmented with external knowledge and tools to interact effectively with users and environments~\cite{Intro5}, and to continually improve themselves via reinforcement learning (RL) with human feedback (RLHF)~\cite{Intro6}.

Early transformer-based PLMs can be grouped into three major architectural categories~\cite{Intro3}: encoder-only, decoder-only, and encoder-decoder models, each shaping the evolution of today's LLMs. While encoder-only models were primarily developed for language understanding tasks~\cite{Intro7}, decoder-only models introduced the generative pre-training paradigm based on next-token prediction~\cite{Intro8}. The latter combines both approaches, unifying NLP tasks under a text-to-text framework and enabling flexible sequence-to-sequence modeling for both comprehension and generation~\cite{Intro9}. Building upon these foundations, the GPT family developed by OpenAI pioneered large-scale self-supervised pre-training~\cite{Intro98}, enabling strong zero- and few-shot performance as well as advanced reasoning and instruction-following capabilities~\cite{Intro8}. Complementing this progress, Meta's open-source LLaMA~\cite{Intro10} and Google's scaled PaLM families~\cite{Intro11} further advanced the field by supporting competitive open-source development and achieving state-of-the-art performance in reasoning and multilingual tasks, respectively. Beyond these, numerous high-performing LLMs, such as Gemini~\cite{Intro110}, DeepSeek~\cite{Intro111,Intro1111}, and LaMDA~\cite{Intro112}, have also achieved strong results and pushed the boundaries of LLM capabilities~\cite{Intro3}.

Using advanced augmentation techniques, LLMs from different families can function as AI agents that perceive their environment, make decisions, and take actions~\cite{Intro12}. Although originally trained as next-token predictors in static settings, agentic behavior requires interacting with dynamic environments, maintaining state across steps, and accessing information beyond the training corpus~\cite{Intro3}. To address limitations such as limited memory, outdated knowledge, hallucinations, and restricted access to external resources, LLM-based agents employ prompt engineering~\cite{Intro13}, retrieval-augmented generation (RAG)~\cite{Intro14}, and tool integration~\cite{Intro15,Intro16}. Prompt engineering structures instructions or reasoning strategies, such as COT prompting, to guide model behavior, while RAG enables querying external knowledge sources. More broadly, tool-augmented LLMs can call application programming interfaces (APIs), perform calculations, search the web, or interact with databases, as seen in systems such as Toolformer~\cite{Intro17}, Gorilla~\cite{Intro18}, and HuggingGPT~\cite{Intro15}, which train models to autonomously select and use tools~\cite{Intro19}. With these capabilities, LLM agents transform static models into interactive, context-aware systems that can reason, verify outcomes, and achieve goals across a wide range of real-world tasks~\cite{Intro20}.

Building on their growing capabilities---especially in agentic settings---LLMs have driven paradigm shifts across the natural and social sciences and accelerated adoption in engineering domains such as software development, hardware design, and materials discovery~\cite{Intro1,Intro21}. In parallel with their expanding role in providing control capabilities and decision support, LLMs are also becoming complex dynamical systems whose behavior must be carefully analyzed and regulated~\cite{Intro22}. Traditional analysis methods provide only partial insight, motivating a shift toward modeling LLMs as evolving systems~\cite{Intro23}. This perspective highlights deep connections with control theory~\cite{Intro233}: both can be represented as dynamical systems, e.g., structured state-space models (SSMs)~\cite{Intro2333}; both steer trajectories toward a desired outcome~\cite{Intro25}; both rely on optimization and feedback mechanisms~\cite{Intro24}; and both prioritize robustness and stability under perturbations~\cite{Intro26}. These parallels underscore why control theory is essential for understanding, regulating, and safely deploying LLMs in real-world engineering settings~\cite{Intro27}.

The complexities of emerging LLM dynamics are even more evident when these models interact with physical systems, such as autonomous systems~\cite{Intro28} and robotics~\cite{Intro29}, where high-level reasoning must align with low-level controllers and tasks demand mathematical precision, physical feasibility, and strict correctness~\cite{Intro30}. Control principles guide LLM outputs, while LLMs support controller design, diagnosis, and adaptation. Despite these advances, the intersection of LLMs and control remains fragmented, with no unified framework. This gap motivates an examination of the conceptual, methodological, and practical connections between LLMs and control theory to develop interpretable, robust, and controllable AI systems for safe integration into engineering and autonomous applications. To clarify this landscape, this study addresses key organizing questions:
\begin{enumerate}[\textbullet]
\item Why is the intersection of LLMs and control important?
\item When did this connection emerge and how has it evolved?
\item Where does control support LLMs? Where do LLMs support control?
\item What are challenges and future trends at this intersection?
\end{enumerate}
\begin{figure}
\centering
\includegraphics[width=\columnwidth]{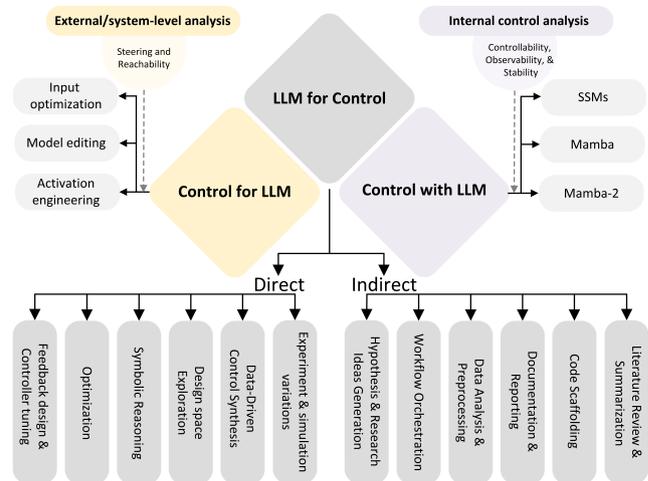}
\caption{Conceptual map of three key intersections between control and LLMs.}
\label{fig_4}
\end{figure}
By addressing these questions, we aim to provide a structured understanding of this rapidly developing field and motivate future work at the convergence of dynamical modeling, control theory, and LLMs. While we have already examined why this intersection matters and will continue to highlight it throughout this study, the insights gained from answering these questions have guided us to the following contributions:
\begin{enumerate}[\textbullet]
\item We trace the historical evolution of the AI-control synergy, from early cybernetics to modern LLM applications, highlighting how these ideas converged.  
\item We categorize interactions between LLMs and control systems into three paradigms (see Fig.~\ref{fig_4}):
\begin{enumerate}[(a)]
\item Indirectly, LLMs support control research and workflow augmentation. Directly, they assist in control design and synthesis, optimization, exploration, reasoning, and guidance of experiments and simulations.
\item Control concepts help LLMs steer their trajectories away from undesired meanings, improving reliability and alignment through external, system-level analyses such as input optimization, model editing, and activation engineering.
\item SSMs enable internal, fundamental control analysis of LLMs through controllability, observability, and stability, thereby ensuring bounded, reliable outputs.
\end{enumerate}
\item We identify challenges and outline future research trends to enhance LLM-driven control, understand LLM behavior, and develop controllable LLMs that are as trustworthy and robust as their electromechanical counterparts.
\end{enumerate}
The proposed interconnection map provides a structured framework for integrating LLMs with control theory, highlighting their potential to enhance research workflows and support control design. Control-theoretic concepts allow LLMs to actively guide their outputs away from undesired behaviors, enhancing alignment and reliability through system-level analysis, while, SSMs enable rigorous internal analysis of the model's dynamics, leveraging controllability, observability, and stability to ensure reliable and bounded outputs. Through this map, we identify current limitations and outline future research directions aimed at developing interpretable and controllable LLMs that are as trustworthy and robust as their electromechanical counterparts, ensuring these models can sustainably support society.

\section{Historical Evolution}
By the mid-20th century, the conceptual foundations of cybernetics and AI were firmly established~\cite{r1}. Norbert Wiener, who was establishing a laboratory at MIT, recognized the significance of McCulloch and Pitts' 1943 neuron model~\cite{r2}, which showed that neural networks (NNs) could, in principle, perform any Turing-computable function (TCF), and invited them to collaborate, sparking the early development of cybernetics---a term still largely unfamiliar at the time. Wiener formalized the field in 1948, identifying feedback as the fundamental mechanism for self-regulating systems~\cite{r3}. These concepts were further disseminated through the interdisciplinary Macy conferences (1946-1953)~\cite{r4}, promoting the application of information and feedback principles across biology, engineering, and the social sciences. This framework guided the design of adaptive systems, exemplified by Ashby's homeostat~\cite{r5}, and inspired early efforts to replicate intelligent behavior in artificial agents. Building on this foundation, AI emerged as a distinct discipline at the 1956 Dartmouth Seminar~\cite{r6}, followed by the development of Rosenblatt's perceptron in 1957~\cite{r7}. The term ``machine learning (ML)'' was later coined in 1959 by Arthur Samuel, an IBM researcher and pioneer in computer gaming and AI~\cite{r8}.

\begin{figure}
\centering
\includegraphics[width=\columnwidth]{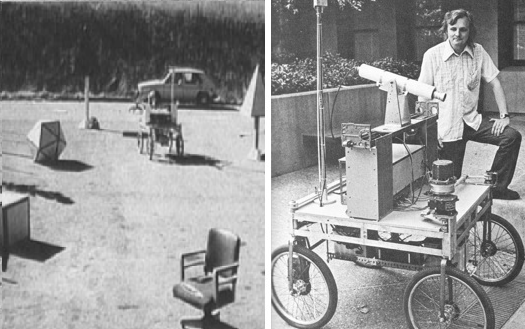}
\caption{Stanford Cart on an obstacle course with a young H. Moravec (1977).}
\label{fig_1}
\end{figure}

\begin{figure}
\centering
\includegraphics[width=\columnwidth]{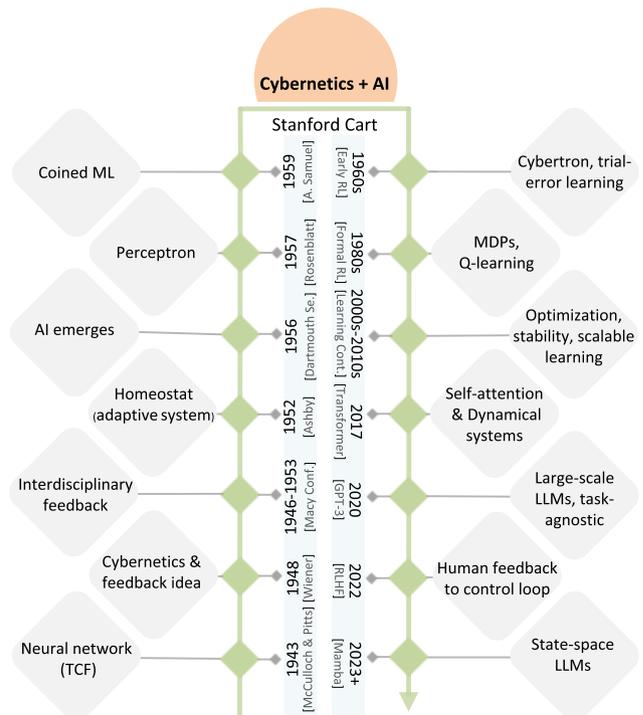}
\caption{When control meets LLMs: A timeline.}
\label{fig_3}
\end{figure}

By the 1960s, cybernetics and AI had become closely intertwined, with feedback-driven principles profoundly influencing various fields, including robotics, economics, and management~\cite{r4}. The Stanford Cart, originally built by James L. Adams to explore the feasibility of controlling a lunar rover from Earth~\cite{r9}, stands as one of the earliest milestones in the \textit{sunrise of cybernetics and AI}, marking the transition from theoretical principles to tangible, feedback-driven machines. Over two decades, graduate students refined the Cart, culminating with Hans Moravec, who added stereo vision to enable autonomous navigation of an obstacle course (see Fig.~\ref{fig_1})~\cite{r10}. Originally developed to simulate lunar rover control, the Cart became an autonomous vehicle capable of following a white line and avoiding obstacles using a TV camera and early computer vision, representing one of the first practical implementations of perception-driven feedback control~\cite{r9}. Building on these foundations, Marvin Minsky's robotic arm in 1968 demonstrated how real-time feedback could be applied to complex motor tasks, balancing strength and precision~\cite{r11}. Together, these developments illustrate how principles of perception, control, and feedback, first explored in early AI and robotics, have evolved into real-world systems that continue to shape technology and human-robot interaction through adaptive, feedback-driven approaches (see Fig.~\ref{fig_3} for the sunrise of cybernetics and AI, and their later convergence leading to the point at which LLMs now meet control theory).

The synergy between control theory and AI became increasingly explicit with the emergence of RL, a term that first appeared in the 1960s when researchers explored trial-and-error methods for adaptive control and pattern recognition~\cite{r12}. In the early 1960s, the Raytheon Company developed an experimental ``learning machine'' known as Cybertron, which used punched-tape memory to analyze sonar signals, electrocardiograms, and speech patterns through a rudimentary form of RL~\cite{r13}. Concurrently, Waltz and Fu (1965)~\cite{r14} and Mendel (1966)~\cite{r15} formalized the term RL to denote systems that adapt their behavior in response to feedback. By the 1980s, these early developments had unified insights from psychology, optimal control, and temporal-difference learning, establishing RL as a distinct discipline. Conceptually, RL is grounded in control-theoretic frameworks such as Markov decision processes (MDPs), paralleling optimal control, where an agent seeks to minimize cumulative cost or maximize reward. A key milestone was Q-learning (Watkins, 1989~\cite{r17}; Watkins and Dayan, 1992~\cite{r18}), which enables agents to learn optimal policies directly from environmental feedback without explicit model knowledge. During the 2000s and 2010s, deep learning transformed AI, while control theory provided crucial insights into optimization dynamics~\cite{r19}, forming a theoretical foundation for large-scale learning systems~\cite{r20}.
\begin{figure}
\centering
\includegraphics[width=\columnwidth]{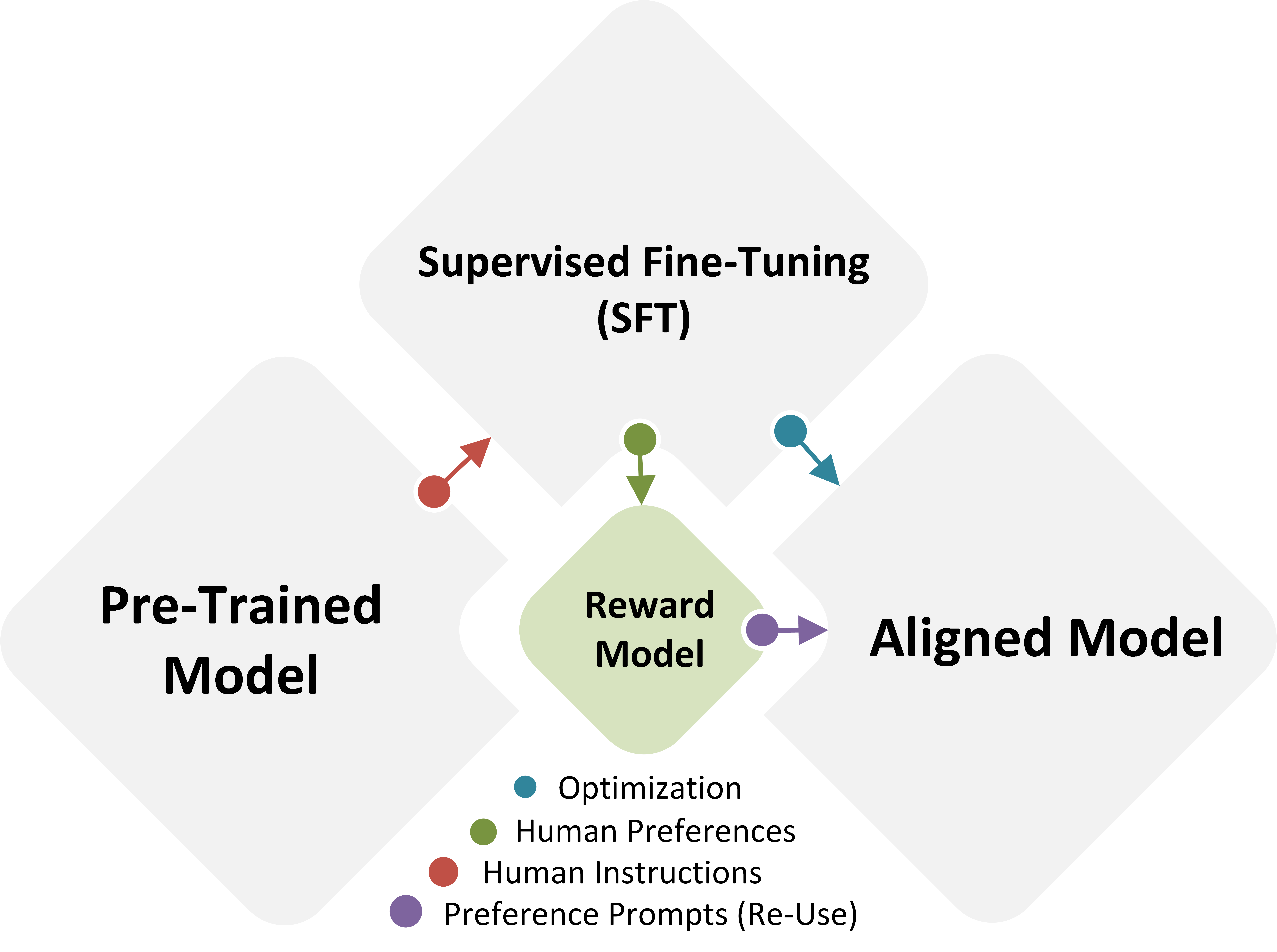}
\caption{The early RLHF pipeline.}
\label{fig_2}
\end{figure}

A major inflection point occurred with the introduction of the Transformer architecture in 2017~\cite{Intro2}. The self-attention mechanism, central to Transformers, has since been interpreted through the lens of dynamical systems, revealing structural parallels to control and signal processing frameworks. LLMs, which began gaining prominence with models such as OpenAI's GPT-3 in 2020~\cite{r20}, demonstrated that scaling up model size could enable broad, task-agnostic language capabilities with minimal task-specific training~\cite{r22}. Subsequently, the advent of RLHF in 2022 embedded explicit feedback loops into LLM fine-tuning pipelines~\cite{r23}. RLHF aligns models with human preferences through a three-step pipeline (see Fig.~\ref{fig_2}): pre-train a language model, train a reward model from human feedback, and optimize the model with RL~\cite{r28}. Essentially, RLHF acts as a control loop, with the model as the system, human feedback as the signal, and RL as the controller, implementing classical control principles like error correction and adaptive regulation. The convergence between control and LLMs has accelerated from 2023 onward. Recent architectures such as Mamba incorporate state-space modeling principles directly into LLM design~\cite{Intro25}, while studies explore controllability, safety, and robustness in generative models. Moreover, LLMs are increasingly applied in real-world control contexts, such as intelligent traffic management~\cite{r25}, transportation~\cite{r26}, and secure automated control system design~\cite{r27}, illustrating a reciprocal flow of ideas between control theory and language modeling. Recent surveys of RLHF emphasize that human preference feedback functions as a closed-loop supervisory signal rather than a peripheral optimization step, positioning feedback as a core mechanism for aligning model behavior~\cite{r28}.

\section{Bridging Control and LLMs}
The link between LLMs and control theory is fundamentally bidirectional, revealing significant opportunities for mutual influence (see Fig.~\ref{fig_4}). On one hand, LLMs can assist in control design by providing both direct technical guidance and broader research support. On the other hand, control-theoretic principles can be used to shape LLMs' behavior, steering their outputs away from undesired meanings and improving reachability~\cite{Intro27,r2777777}. Besides, recent studies have overreached further by treating LLMs as dynamical state-space systems, enabling researchers to analyze their internal representations in terms of controllability and observability, apply stability analysis, and link them to external controllers that monitor outputs and provide corrective signals, analogous to physical control systems. The following subsections detail these complementary directions:

\subsection{LLM for Control}
LLMs are emerging as powerful tools in control engineering, supporting both the research and the design aspects of dynamical systems control~\cite{DSEddc2,DSEddc3,DSEddc4}. Their contributions can be categorized along two complementary dimensions: \textbf{indirect} applications, where LLMs enhance research workflows, and \textbf{direct} applications, where they assist with the technical aspects of control system development and operation. Together, these modes illustrate how LLMs integrate into the control ecosystem---not as replacements, but as augmentation mechanisms that extend human capability.

\subsubsection{Indirect (Research and Workflow Support)}
LLMs are transforming control engineering research by acting as intelligent assistants that automate repetitive yet essential tasks. Beyond literature reviews and summarization~\cite{r299}, they can scaffold simulation code~\cite{r29,DSEres2}, extract and organize mathematical formulations, and convert raw experimental logs into coherent technical reports~\cite{r2999}. They can also preprocess complex, heterogeneous datasets, assessing data quality, structure, and key properties such as persistent excitation and data informativity~\cite{r29999}, supporting the development of reliable data-driven controllers~\cite{r299999}. By automating core control-research tasks, including data cleaning and processing, experiment design and management, simulation setup, and results visualization, LLMs streamline the orchestration of complex workflows. They augment the intellectual capabilities of control engineers, enabling advanced reasoning, novel hypotheses generation, and innovative control-theoretic investigations~\cite{r30}. This augmentation not only accelerates the research cycle but also leads to more substantial and impactful scientific and engineering outcomes. To formalize this augmentation, LLM-enhanced research workflows can be structured into three progressive capability levels, summarized in Table~\ref{tab:llm_levels}:
\begin{enumerate}[\textbullet]
\item Level 0 treats prompting as a utility function, where structured system messages perform specific tasks such as argument construction or hypothesis generation. As shown in Listing~\ref{lst:level0}, the LLM receives a system message defining its role as a control-systems expert and a prompt containing the claim to evaluate, producing reproducible arguments for rapid hypothesis formulation.
\item Level 1 extends LLM capabilities by producing structured outputs using frameworks like Pydantic~\cite{r300}, capturing hypotheses, assumptions, or experimental evidence as validated data primitives for reproducible comparison across plants, objectives, constraints, and stability guarantees. Listing~\ref{lst:level1} shows the LLM evaluating a research paper in a consistent, semantically validated format suitable for automated aggregation and controller comparison.
\item Level 2 represents fully agentic workflows, where LLMs autonomously perform multi-step research tasks, including searching, filtering, summarizing, and drafting literature reviews or experimental reports. Listing~\ref{lst:level2} demonstrates how, for example, the GPTResearcher agent orchestrates a complete research pipeline~\cite{r301}, producing structured and verifiable reports that would otherwise require weeks of manual effort.
\end{enumerate}

\begin{table*}
\caption{Levels of LLM-augmented control research workflows$^*$.}\label{tab:llm_levels}
\resizebox{\textwidth}{!}{\selectfont
\begin{tabular*}{\tblwidth}{l p{2cm} p{6.5cm} p{6.5cm} }
\toprule
\textbf{Level} & \textbf{Concept} & \textbf{Technical Functionality} & \textbf{Example}\\
\toprule
L-0 & 
Utility-based reasoning & (1) Structured messages; (2) Atomic reasoning (argument construction, hypothesis generation) & Evaluating a control-theoretic claim through a structured prompt (e.g., assessing if a given controller improves stability) (Listing~\ref{lst:level0}) \\\midrule 
L-1 & 
Semantic data structures & (1) Integration with structured data or schema validation frameworks; (2) Validated schema for hypotheses, assumptions, and evidence & 
Structured outputs to assess paper relevance via Pydantic, enabling reproducible evaluation of hypotheses \& controller comparisons (Listing~\ref{lst:level1}) \\\midrule 
L-2 & 
Autonomous multi-step orchestration & (1) Sequential LLM for literature search, filtering, summarizing, \& report drafting; (2) Integration with external APIs \& structured outputs & 
Executing autonomous research pipelines with an agentic LLM, like GPTResearcher, for literature search \& report generation (Listing~\ref{lst:level2})\\
\bottomrule
\multicolumn{4}{l}{*OpenAI, Pydantic, and GPTResearcher are used here as illustrative examples; equivalent LLMs, structured data frameworks, or agentic}\\
\multicolumn{4}{l}{ tools could be substituted at each workflow level.}
\end{tabular*}}
\end{table*}

\begin{lstlisting}[style=pythonstyle,
    caption={Prompting for verifiable research tasks (L-0).},
    label={lst:level0}]
from openai import OpenAI
client = OpenAI()

SYS_MSG = """You are a control-systems expert. Argue for or against the given claim."""
prompt = "Claim: model predictive control (MPC) improves power grid stability."

resp = client.chat.completions.create(
  model="gpt-4o-mini",
  msg=[{"role": "system", "content": SYS_MSG},
       {"role": "user", "content": prompt}])

print(resp.choices[0].message.content)
\end{lstlisting}
\begin{lstlisting}[style=pythonstyle,
    caption={Structured outputs using Pydantic (L-1).},
    label={lst:level1},
    basicstyle=\footnotesize\ttfamily
]
from openai import OpenAI
from pydantic import BaseModel, Field
from typing import Literal
client = OpenAI()

class PaperRelevance(BaseModel):
    relevancy: Literal['yes', 'no']
    justification: str = Field(description="One-sentence justification")

SYS_MSG = """Assess paper relevance as control engr. Return 'yes' or 'no' with justification."""

def assess_relevance(topic, summary):
    resp = client.beta.chat.completions.parse(
        model="gpt-4o-mini",
        msg=[{"role": "system", "content": SYS_MSG},
            {"role": "user", "content": f"Topic: {topic}\nPaper: {summary}"}],
        response_format=PaperRelevance)
    return resp.choices[0].message.parsed

topic = "Distrib. MPC for grid freq. regulation."
summary = "Hierarch. MPC for distrib. grid control."
print(assess_relevance(topic, summary))
\end{lstlisting}
\begin{lstlisting}[style=pythonstyle,
    caption={Agentic Workflow for Autonomous Research (L-2).},
    label={lst:level2},
    basicstyle=\footnotesize\ttfamily
]
from gpt_researcher import GPTResearcher
import asyncio

async def autonomous_review():
    researcher = GPTResearcher(query="Distributed MPC for power grid stability")
    results = await researcher.conduct_research()
    report = await researcher.write_report()
    print(report)

asyncio.run(autonomous_review())
\end{lstlisting}

This hierarchy, supported by a broader ecosystem of chatbots, hybrid search models, and academic LLM platforms, demonstrates how LLMs can augment control engineering workflows. Together, they enable reproducible, verifiable, and efficient research grounded in both human expertise and LLM-assisted analysis. Beyond the hierarchy itself, a growing set of GenAI tools supports specific stages of the research process~\cite{r302}. General-purpose chatbots~\cite{r303}, e.g., ChatGPT, Claude, assist with brainstorming, keyword generation, and search refinement, while chatbot-search hybrids such as Copilot and Perplexity use real-time internet access to deliver updated, context-aware insights~\cite{r304}. Academic-oriented tools, e.g., Scite.ai, Consensus, Scopus AI, Elicit, Research Rabbit, Connected Papers, and Litmaps, specialize in literature discovery, synthesis, and citation mapping. Together, these tools enable end-to-end automation of domain research tasks, from question formulation and evidence retrieval to structured synthesis and validation, aligning with Level 2 agentic workflow principles.

\subsubsection{Direct (Control System Designs)}
Beyond supporting research workflows, LLMs have now demonstrated the ability to directly influence control system design, actively stepping into the design loop itself. They can participate in controller synthesis, parameter tuning, and experimental design. Key functional roles include:
\begin{enumerate}[\textbullet]
    \item \textbf{Feedback design:} Suggesting controller parameters (e.g., linear quadratic regulator (LQR) weights or proportional-integral-derivative (PID) gains) to meet specifications.
    \item \textbf{Optimization:} Guiding control decisions by planning sequences of actions or reasoning trajectories that approximately minimize cost functions.
    \item \textbf{Symbolic reasoning:} Assisting in converting high-level descriptions into formal representations, such as analytical expressions, logical rules, or control-theoretic constructs (e.g., candidate Lyapunov or cost functions).
    \item \textbf{Design space exploration:} Generating and evaluating alternative system architectures or control strategies, and systematically analyzing trade-offs to guide design choices.
    \item \textbf{Data-driven modeling and analysis:} Leveraging knowledge extracted from datasets, LLM-based agents autonomously construct data-driven models for complex engineering systems, achieving predictive accuracy and analytical performance comparable to state-of-the-art, human-developed approaches.
    \item \textbf{Experiment \& simulation variations:} Recommending adjustments to experimental setups or simulation configurations that directly guide research and design decisions.
\end{enumerate}
In essence, these roles enable LLMs to contribute directly to control system design, complementing human expertise and accelerating innovation, \emph{without supplanting critical judgment}. Such contributions can exploit the LLM's pre-trained knowledge through zero-shot prompting or be augmented via few-shot prompting, which directs the model to generate task-specific proposals and iteratively refine them to meet performance, safety, and precision requirements~\cite{Intro1,Intro3}:
\begin{enumerate}[\textbullet]
\item \textbf{Zero-shot prompting:} Generate outputs by ``following task instructions'' without task-specific examples, generalizing from prior knowledge to novel system dynamics or design scenarios, and supporting rapid exploration of new architectures or previously unseen environments.
\item \textbf{Few-shot prompting:} Uses ``in-context learning (ICL)'' with a small set of examples to refine control strategies, satisfy precise specifications, and respect safety or performance constraints, particularly in high-accuracy tasks.
\end{enumerate}
While in zero-shot settings the LLM generates initial hypotheses and faces the challenge of generalizing to novel tasks or partners~\cite{Intro1}, in few-shot it iteratively refines its outputs to enhance performance and feasibility. To support multi-step reasoning in control tasks, CoT prompting encourages LLMs to produce intermediate reasoning steps, yielding structured and interpretable solutions even under zero- or few-shot setups~\cite{Dr1}. In what follows, we examine the above-mentioned key functionalities in detail and provide representative examples that illustrate how LLMs can directly participate in the control design process under different types of prompting\footnote{LLM integration in control can be classified as: (1) fine-tuning for specific tasks, (2) pairing with trainable components, or (3) direct use of pre-trained models~\cite{DSEfeed7}, with zero-shot, few-shot, and CoT prompting generally employed in the third category.}.

\paragraph{LLM-Based Feedback Design (Controller Tuning):}
Recent studies indicate that LLMs have acquired knowledge relevant to control engineering and can address textbook-level control questions~\cite{DSEddc3}. However, they remain unreliable for practical control design, as they can make reasoning errors and struggle with the subtle performance–robustness trade-offs inherent in real-world systems. Nevertheless, LLMs show promise in supporting control system design, particularly in automating iterative or tedious tasks such as controller tuning~\cite{DSEfeed1,DSEfeed2}. Classical feedback controllers are widely used in industrial applications, with PID control and loop-shaping being the most prevalent due to their simplicity and ease of implementation. Over the years, numerous PID and loop-shaping tuning methods have been developed~\cite{DSEfeed22,DSEfeed222,DSEfeed2222}. Despite these advancements, tuning still heavily relies on human expertise and manual trial-and-error to identify parameters that satisfy design requirements~\cite{DSEfeed3}.

Recently, LLMs have demonstrated the ability to bridge gaps in control system design by integrating with human expertise to translate performance requirements into actionable parameters, generate candidate gain sets, and iteratively refine them using structured prompting and optimization~\cite{Intro30,DSEfeed1,DSEfeed2,DSEfeed4,DSEfeed5,DSEfeed33}. Frameworks such as ControlAgent~\cite{DSEfeed1} and AgenticControl~\cite{DSEfeed2} employ agent-based approaches where LLMs collaborate with computational modules to propose, evaluate, and optimize controller configurations. Similarly, SmartControl uses LLM-guided optimization to determine PID gains from natural language specifications, producing stable and efficient closed-loop responses~\cite{DSEfeed4}. Controller tuning falls under direct use of pre-trained models, with LLMs and ICL generating or refining gains without retraining~\cite{DSEfeed7}. In this paradigm, LLMs support feedback control design in both offline (out-of-the-loop) workflows, where gains are proposed and verified externally, and online (in-the-loop) workflows, where iterative refinement is guided by performance feedback (see Fig.~\ref{fig_5}):

\begin{figure}
\centering
\includegraphics[width=0.95\columnwidth]{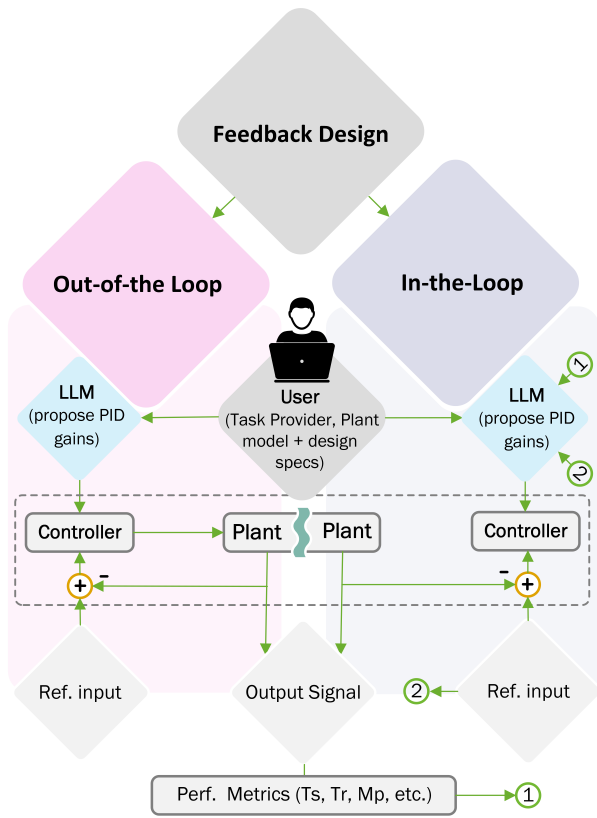}
\caption{LLM-assisted controller tuning (out-of-the-loop and in-the-loop).}
\label{fig_5}
\end{figure}

\begin{figure}
\centering
\includegraphics[width=0.95\columnwidth]{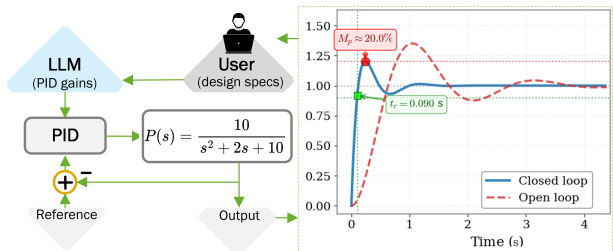}
\caption{Step response of the plant in Listing~\ref{lst:offline_pid} under the PID controller with gains obtained via offline LLM-based tuning.}
\label{fig6}
\end{figure}

\textbf{Out-of-the-Loop (Offline) Tuning:} In this tuning method, the LLM proposes controller parameters without interacting directly with the closed-loop system, while a human designer or simulation environment evaluates the results. This mirrors traditional practice in which PID gains are first proposed analytically and then validated through simulation. Pre-trained LLMs are effective in this setting: given high-level performance requirements (e.g., rise time $T_r$, settling time $T_s$, overshoot limits $M_p$, etc.), they can quickly generate multiple candidate gain sets, reducing manual trial-and-error while maintaining human oversight. Verification occurs in simulation, with iterative refinement requested as needed. To illustrate the workflow, the Python script in Listing~\ref{lst:offline_pid} queries DeepSeek-R1 via the Ollama API, providing the plant transfer function and desired closed-loop criteria. The LLM returns PID gains in parallel form ($K_p=10.25$, $K_i=48.6$, $K_d=1.4$), which yield a rise time under 1 second and approximately 20\% overshoot (see Fig.~\ref{fig6}). These results demonstrate that LLM-guided offline tuning offers analytically grounded initial designs while preserving human oversight.

\begin{lstlisting}[style=pythonstyle,
    caption={Offline PID tuning using DeepSeek-R1.},
    label={lst:offline_pid},
    basicstyle=\footnotesize\ttfamily
]
import ollama

Specify model
desiredModel = 'deepseek-r1:32b'
Define control design question

questionToAsk = ('Design a PID controller in the parallel form for the open-loop plant '
'given by 10/(s^2 + 2*s + 10). The closed-loop system should have a '
'rise time less than 1 second and an overshoot less than 20 percent.')

Query the LLM

response = ollama.chat(model=desiredModel,
msg=[{'role': 'user', 'content': questionToAsk}],)

Extract and display response

OllamaResponse = response["message"]["content"]
print(OllamaResponse)

Save output

with open("OutputOllama.text", "w", encoding="utf-8") as f:
f.write(OllamaResponse)
\end{lstlisting}

\textbf{In-of-the-Loop (Online) Tuning:} Recent advances have formalized the paradigm of iterative, in-the-loop tuning through LLM-guided approaches, establishing them as powerful tools for automated controller design~\cite{DSEfeed1,DSEfeed2,DSEfeed4}. In this paradigm, LLMs actively participate in the real-time adjustment of controller parameters by observing system responses and iteratively refining control actions~\cite{DSEfeed6}, which is useful for adaptive systems, where uncertainties or time-varying dynamics make offline tuning insufficient. As shown in Fig.~\ref{fig_5}, in addition to the Controller, which receives the error signal and computes the control input using the current PID parameters, and the Plant, which produces the system output, the basic in-the-loop tuning workflow includes:
\begin{enumerate}[\textbullet]
\item Performance Metrics Unit: Monitors the system response and computes performance metrics.
\item LLM Tuner: Receives the current PID parameters and performance metrics and proposes updated parameters to iteratively improve the closed-loop response.
\end{enumerate}
This feedback-driven loop allows the LLM to continuously refine controller gains, ensuring convergence toward the desired performance while maintaining stability. The framework interfaces with LLMs via APIs, supporting GPT-based models or locally hosted models (DeepSeek, LLaMA, Mistral), which enable real-time adaptability through ICL.

A prominent example is ControlAgent~\cite{DSEfeed1}, a multi-agent framework in which a central task distributor coordinates specialized LLM agents that iteratively generate candidate controllers, evaluated by a Python computation agent. Following a loop-shaping methodology, each iteration refines parameters such as proportional and integral gains based on the closed-loop transfer function. Evaluated on the ControlEval benchmark, which covers stable and unstable, first- and second-order, delayed, and higher-order systems, ControlAgent achieved near-perfect success rates. The LLM input prompt $\mathcal{P}$ consists of (i) design instruction $\mathcal{E}_{\text{spec}}$, encoding domain expertise and guiding performance–robustness trade-offs using PID or loop-shaping methods; (ii) user requirements $\mathcal{U}$ specifying desired performance; (iii) memory $\mathcal{M}$ and feedback $\mathcal{F}$ storing prior designs and evaluations while highlighting deficiencies; and (iv) response instruction $R$ defining the output format for efficient extraction and iterative incorporation (see Algorithm~\ref{alg1}). Upon receiving task requirements from the central agent $\mathcal{A}_c$, the task-specific LLM agent $\mathcal{A}_{\text{spec}}$ generates and stores new controller designs based on instructions, prior failures, and feedback, which are evaluated by a Python agent $\mathcal{A}_p$; the process iterates with performance-based feedback until requirements are met or the iteration limit is reached.

\begin{algorithm}
\vspace*{0.4mm}
\caption{In-the-loop controller design process.}\label{alg1}
\begin{algorithmic}[1]
\State \textbf{Input:} User requirements $\mathcal{U}$, maximum iterations $N$
\State \textbf{Output:} Designed controller $C$
\State Initialize structured memory buffer $\mathcal{M} \leftarrow \varnothing$
\State Initialize feedback $\mathcal{F}_0 = \{\}$
\State Task assignment: $\mathcal{A}_c$ assigns task to $\mathcal{A}_{\text{spec}}$ based on $\mathcal{U}$
\For{$k = 1$ to $N_{\text{max}}$}
    \State Generate $\mathcal{P}_k \leftarrow \text{GenPrompt}(\mathcal{E}_{\text{spec}}, \mathcal{U}, \mathcal{M}, \mathcal{F}_{k-1})$
    \State Generate controller $C_k \leftarrow \mathcal{A}_{\text{spec}}(\mathcal{P}_k)$
    \State Update memory $\mathcal{M} \leftarrow \mathcal{M} \cup \{C_k\}$
    \State Python agent $\mathcal{A}_p$ evaluates $C_k$ and computes performance $P_k$
    \If{$P_k$ satisfies $R$}
        \State \textbf{Return:} Successfully designed controller $C_k$
    \EndIf
    \State Generate feedback $\mathcal{F}_k \leftarrow \text{GenFeedback}(P_k, R)$
\EndFor
\State \textbf{Return:} No successful controller found
\end{algorithmic} 
\end{algorithm}

Despite achieving up to 100\% success rate on first- and second-order stable systems with minimal iterations and outperforming all LLM-based and traditional PID tuning baselines, ControlAgent remains limited to linear time-invariant (LTI) systems and conventional controllers, highlighting the need for future extensions to nonlinear, adaptive, and robust control scenarios. Nevertheless, it has served as the foundation for several related approaches. AgenticControl~\cite{DSEfeed2} employs specialized LLM agents in an actor-critic-juror loop to propose, evaluate, and refine parameters for PID and full-state feedback controllers, achieving robust performance across (non)linear systems. SmartControl~\cite{DSEfeed4} combines LLMs with population-based optimization to translate natural-language specifications into implementable PID designs, providing interactive feedback and modular support for extensions such as fractional PID and nonlinear control. A major extension is LLM-guided adaptive PID controllers, where an explicit plant model and optional initial controller structure enable the LLM to synthesize adaptive control laws, iteratively refined from observed system responses~\cite{Intro30,DSEfeed5,DSEfeed6}. These results demonstrate that LLMs can serve as intelligent, adaptive assistants, capable of refining controllers while preserving safety and enabling faster, robust, and flexible tuning.

\paragraph{LLM-Based Optimization:}
Methods such as CoT prompting~\cite{Intro4}, self-consistency~\cite{DSEopt2}, ReAct~\cite{Intro6}, and agentic frameworks~\cite{DSEopt4,DSEopt5} demonstrate that guiding LLM outputs with structured prompts improves performance on complex, multi-step tasks. These approaches suggest that LLMs can implicitly evaluate and optimize sequences of actions when appropriately prompted, effectively acting as approximate planners. In robotics and embodied agents, systems such as SayCan~\cite{DSEopt6} and Voyager~\cite{DSEopt7} show that LLMs can generate action plans in complex environments. Evaluating these plans with cost or reward functions, analogous to Monte Carlo tree search-style approaches~\cite{DSEopt8,DSEopt9}, further improves performance. This viewpoint naturally aligns with MPC, a well-established framework in control theory where an agent generates sequences of actions over a finite horizon to minimize a predefined cost function~\cite{DSEopt10}. In MPC, the cost function typically combines a task-specific objective, such as distance to a goal state—with regularization terms that penalize aggressive or complex control actions. Similarly, LLMs can be prompted to propose multiple candidate action sequences, which are then evaluated with respect to an objective function to select the best sequence for execution.

Specifically, an agent navigates a state space $s_t$ by selecting actions $a_t$. The state evolves according to $s_{t+1} = f(s_t, a_t, \varepsilon_t)$, where $\varepsilon_t$ represents a noise or disturbance term. The objective is to find a sequence of actions $a_t, \dots, a_{t+H}$ over a horizon $H$ that minimizes a cost function:
\begin{equation*}
\begin{aligned}
\hat a_t, \dots, \hat a_{t+H} &= g(s_t) \\
:&= \arg \min_{a_t, \dots, a_{t+H}} C(\{s_{j}\}_{j=t}^{t+H},\{a_{j}\}_{j=t}^{t+H}),
\end{aligned}
\end{equation*}
where $C$ includes task-specific objectives and regularization terms. As described in Algorithm~\ref{alg2}, an LLM operates on a sequence of tokens $q_t \in \{1, \dots, M\}^L$ and outputs a probability distribution over the next token. The prompt incorporates state information: $q_t = P(s_t, q_{t-1})$. Repeated evaluation produces a token sequence $\hat Y \in \{1, \dots, M\}^T$, i.e., $\hat Y = F(q_t)$. Mapping tokens to actions yields
\begin{equation*}
\begin{aligned}
\hat a_t, \dots, \hat a_{t+H} = \varphi(\hat Y) = \varphi(F(P(s_t, q_{t-1}))) \approx g(s_t).
\end{aligned}
\end{equation*}
To improve performance, multiple plans $A^i_t = \{a^i_t, \dots, a^i_{t+H}\}$ are sampled from the LLM, simulated to obtain trajectories $S^i_t = \{s^i_t, \dots, s^i_{t+H}\}$, and evaluated using the cost function. The optimal plan is selected as $\hat{A}_t = \arg \min C(S^i_t, A^i_t)$, for $i=1,\dots,K$. Up to $H$ actions of $\hat A_t$ are executed, after which the process repeats from the updated state $s_{t+1}$. As an illustrative prompt template, an LLM can generate $K$ candidate control sequences for a mass-spring system with specified dynamics and objectives (see Listing~\ref{lst:LLM-MPC}). Each sequence is a list of length $H$ with values between $0$ and $20$, and all sequences are returned as a Python dictionary. The sequence with the lowest cost is then selected for execution.

\begin{algorithm}
\vspace*{0.4mm}
\caption{LLM as MPC plan sampler.}\label{alg2}
\begin{algorithmic}[1]
\State \textbf{Input:} $s_t$, $q_{t-1}$, $H$, $K$
\State \textbf{Output:} selected action sequence $\hat{A}_t = \{\hat{a}_t, ..., \hat{a}_{t+H}\}$
\State Initialize current prompt: $q_t \gets P(s_t, q_{t-1})$
\For{$i = 1$ to $K$}
    \State Sample token sequence from LLM: $\hat{Y}^{(i)} \gets F(q_t)$
    \State Map token-to-action sequence: $A_t^{(i)} \gets \phi(\hat{Y}^{(i)})$
    \State Simulate system: $S_t^{(i)} \gets \{s_t^{(i)}, ..., s_{t+H}^{(i)}\}$ \\
           where $s_{t+h+1}^{(i)} = f(s_{t+h}^{(i)}, a_{t+h}^{(i)}, \epsilon_{t+h})$
    \State Evaluate plan cost: $J^{(i)} \gets C(S_t^{(i)}, A_t^{(i)})$
\EndFor
\State Select best plan: $\hat{A}_t \gets \arg \min_{i=1,...,K} J^{(i)}$
\State Execute first $H$ actions of $\hat{A}_t$ on the system
\State Observe new state $s_{t+H+1}$ and repeat from step 1
\State \textbf{Return:} $\hat{A}_t$
\end{algorithmic} 
\end{algorithm}

\begin{lstlisting}[style=pythonstyle,
    caption={Offline PID tuning using DeepSeek-R1.},
    label={lst:LLM-MPC},
    basicstyle=\footnotesize\ttfamily
]
from openai import OpenAI

def query_llm_for_plans(x_init, v_init, H, K, dt, m, k_spring, x_goal):
    prompt = f""" Given:
    - A mass-spring system: position x & velocity v
    - Parameters: m={m}, k_s={k_spring}, dt={dt}
    - Dynamics: x_(k+1)= x_k + dt*v_k,
                v_(k+1)= v_k + (dt/m)*(u_k-k_s*x_k)
    - Current state: x={x_init}, v={v_init}
    - Goal state: x_goal={x_goal}
    - Horizon: H={H}
    - Current spring force: {-k_s*x_init}
    - Cost: C_{t+H} = Q_{x,v}({x,v}_{t+H}-{x,v}^*)^2 + \sum_{k=t}^{t+H} Q_u u_k^2

    You control the spring via u = [u_0, ..., u_{H-1}] to reach the goal. Propose {K} candidate control sequences, each a list of length H with values between 0 and 20. Return them as a Python dictionary with keys "seq_1", ..., "seq_{K}", where each value is the corresponding control sequence. Example format:
    {{"seq_1": [u_0, u_1, ..., u_{H-1}],
        "seq_2": [u_0, u_1, ..., u_{H-1}],
        ...}}
    Do not include code blocks or extra commentary; return only the dictionary. """
    return prompt
\end{lstlisting}

\paragraph{LLM-Assisted Symbolic Reasoning:}
LLMs are achieving human-level performance across a wide range of tasks, yet their reasoning capabilities, which include factuality, compositionality, rule discovery, and alignment with real-world knowledge, remain an active area of research~\cite{DSEres1}. Mathematics is widely regarded as one of the most demanding forms of reasoning~\cite{DSEres4}. Consequently, the ability to solve differential equations and perform scientific computing tasks represents a significant milestone for LLMs~\cite{DSEres3,DSEres33}. Most studies focus on problems with known solutions~\cite{DSEres5,DSEres6,DSEres7,DSEres8}, while relatively few address open-ended challenges, such as combinatorial optimization or graph-theoretic problems~\cite{DSEres9,DSEres10}. This is partly due to the difficulty of generating training data for open problems and the need for substantial expert preparation. Many tasks in control research, such as deriving stability conditions, formulating symbolic controller laws, constructing system representations, and translating high-level descriptions into formal control-theoretic specifications, require rigorous symbolic and mathematical reasoning. These demands closely parallel those of advanced mathematics, making control theory an ideal domain to study and evaluate the reasoning capabilities of LLMs.

A notable example is Meta's recent work on discovering global Lyapunov functions~\cite{DSEres11}, where the goal was to find an analytical function $V(x)$ guaranteeing the global stability of a dynamical system $\dot{x} = f(x)$. Meta trained sequence-to-sequence transformers on synthetic datasets pairing stable systems with Lyapunov functions, created via a \textit{backward generation} method, sampling a candidate $V$ and then constructing $f(x)$ to satisfy $V(0) = 0$, $V(x) > 0$, for $x \neq 0$, $\lim_{\|x\| \to \infty} V(x) = \infty$, and $\dot{V}(x) \le 0$, and via \textit{sum-of-squares} methods for polynomial systems.  The model mapped system definitions to symbolic Lyapunov functions in pre-order tree notation, achieving 99\% accuracy on in-distribution polynomials, 73–75\% on out-of-distribution tests, and up to 12.7\% success on non-polynomial systems. This approach illustrates LLMs' roles in deriving symbolic objects, validating stability analytically, and guiding human or algorithmic controller synthesis. Despite these successes, LLMs face challenges in reasoning about physical and dynamical systems, as they are trained solely on text. However, performance improves when complex tasks are decomposed, as in~\cite{DSEfeed1}, where few-shot LLM agents synthesized controllers from natural language specifications.

\paragraph{Design Space Exploration (Robot Planing):}
Traditional planning and control algorithms offer strong theoretical guarantees but require substantial expertise and often struggle to adapt or scale in complex, dynamic environments. LLMs provide a promising alternative by enabling automated reasoning over task specifications, environmental constraints, and system dynamics. Few-shot prompting allows LLMs to autoregressively predict low-level robotic actions without task-specific fine-tuning~\cite{DSEr1}, while integration with predictive controllers enables dynamic adaptation of unmanned aerial vehicle (UAV) dynamics and low-level parameters~\cite{DSEr2,DSEr3}. Hierarchical prompting facilitates autonomous generation of control code from natural language instructions~\cite{DSEr4}, as seen in systems such as Text2Motion~\cite{DSEr5}, ProgPrompt~\cite{DSEr6}, and EUREKA~\cite{DSEr7}, which translate instructions into executable plans or reward functions using iterative feedback, structured programmatic prompting, and pretrained skill modules~\cite{DSEopt6,DSEr6,DSEr9,DSEr10}. Beyond single-robot tasks, LLMs have been combined with motion planning for multi-object rearrangement~\cite{DSEr11}, refining goals, constraints, and reward structures for RL~\cite{DSEr12,DSEr13,DSEr14}, and designing curricula for efficient policy learning by decomposing tasks into subtasks and generating executable training code~\cite{DSEr15}. They have also been applied to language-driven zero-shot object navigation, guiding exploration and sequential decision-making using commonsense reasoning and pretrained vision-language models~\cite{DSEr16}, with closed-loop linguistic feedback further improving planning and reasoning~\cite{DSEr17}. These approaches demonstrate that LLMs can explore design spaces, iteratively refine strategies based on performance feedback, and enable more adaptive robotic systems~\cite{DSEr5}.

\begin{figure}
\centering
\includegraphics[width=\columnwidth]{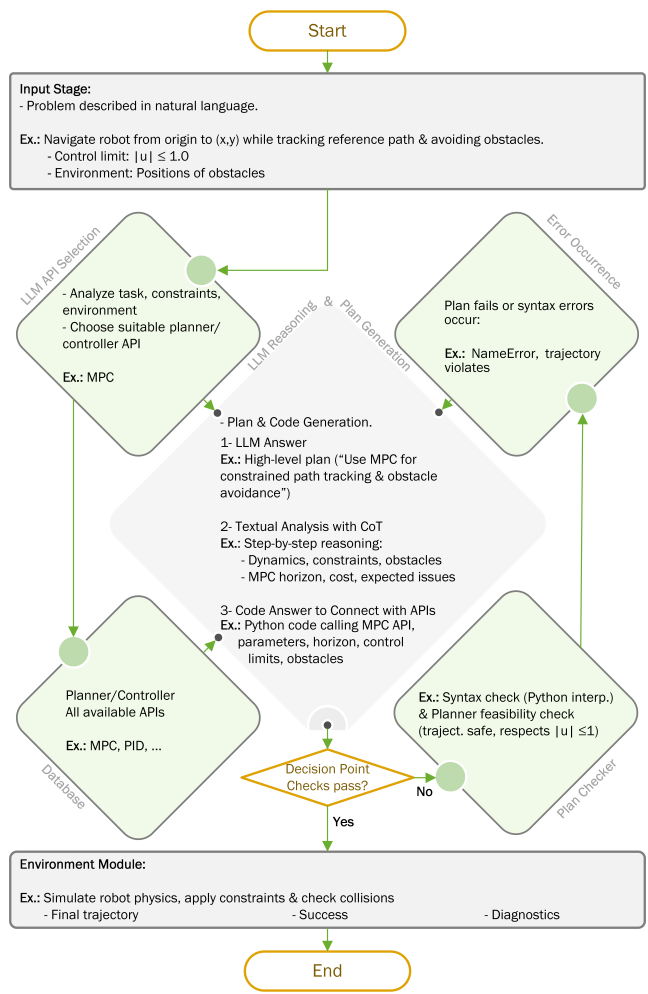}
\caption{Architecture of the LLM-based strategy decision framework integrating task analysis, planner selection, and iterative feedback, illustrated with a robot navigation example.}
\label{fig_8}
\end{figure}

Recent advances in agentic LLM architectures have formalized the internal organization of autonomous reasoning systems, identifying key modules, such as perception, memory, and execution, that enable adaptive decision-making and closed-loop behavior~\cite{DSEr19}. These modules allow LLMs to perceive task contexts, plan actions using techniques such as CoT reasoning, retain relevant knowledge for continuous improvement, and execute actions in real or simulated environments. Frameworks such as AuDeRe demonstrate practical applications~\cite{Intro29}, interpreting high-level natural language task descriptions and selecting control algorithms (e.g., MPC, LQR) from a database of APIs rather than directly generating trajectories or code. As shown in Fig.~\ref{fig_8}, the framework systematically selects, integrates, and iteratively refines strategies by analyzing task objectives, environmental constraints, and system dynamics, retrieving relevant API definitions, input/output specifications, and integration requirements to construct executable plans. For example, in a robot navigation task, AuDeRe identifies suitable algorithms, configures parameters, and refines solutions through closed-loop feedback when encountering planning errors, syntax issues, or constraint violations; a trajectory-tracking scenario with a nonlinear Dubins car illustrates iterative MPC adaptation for collision-free navigation. Beyond robotics, similar principles apply to autonomous driving: DriveLLM integrates LLMs with traditional driving stacks to enable commonsense reasoning, hazard prediction, instruction interpretation, and adaptation through cyber-physical feedback~\cite{DSEr21}. These examples show how iterative reasoning, modular integration, and closed-loop feedback enable LLMs to explore the design space, generating, evaluating, and refining candidate solutions to extend planning and control to complex and safety-critical systems.

\paragraph{Data-Driven Modeling and Analysis:}
Modern science and engineering increasingly rely on data-driven modeling to extract insight, enable accurate prediction, and support analysis and control of complex systems as high-fidelity simulations and experiments generate vast datasets~\cite{DSEddcnew1}. AI and ML methods, particularly deep NNs, have shown strong ability to model nonlinear, multiscale behaviors in engineering applications. However, building reliable data-driven models remains labor-intensive and case-specific, limiting scalability. Recent advances in LLMs offer a promising path to automate and streamline this process. Beyond natural language understanding, modern LLMs function as general-purpose reasoning engines that, when embedded in agentic frameworks and coupled with computational tools, can autonomously perform multi-step analytical workflows~\cite{DSEddcnew11}. By integrating reasoning, coding, and data interaction, LLM-based agents have the potential to automate and streamline end-to-end data-driven modeling and analysis, enabling efficient development of high-quality models for complex engineering systems and advancing data-informed decision-making and control~\cite{DSEddcnew1}.

Power electronic (PE)-GPT exemplifies this paradigm as a physics-informed LLM for modulation design in power converters~\cite{DSEddc6}, later extended in~\cite{DSEddc6new} with integrated design, simulation verification, and model fine-tuning workflows. Through text-based interaction, PE-GPT translates user specifications into optimized modulation strategies by combining metaheuristic optimization with hierarchical physics-informed NNs (ModNet and CirNet), achieving rapid and accurate design synthesis, as demonstrated by a dual active bridge converter with a 63.2\% accuracy improvement and validated zero-voltage switching. Complementing this design-oriented framework, LLMs have also been applied to data-driven control for distribution voltage regulation. Motivated by recent advances in few-shot ICL~\cite{DSEddc66}, an LLM-based inverter control scheme constructs a topology-adaptive surrogate power-flow model from limited data and embeds it into an MPC framework, enabling real-time voltage regulation under frequent topology changes without explicit grid models~\cite{DSEddc7}. In this framework, the distribution grid is modeled with photovoltaic inverters located at nodes $\mathcal{H} \subseteq \mathcal{N}$, where each PV unit $i$ has active and reactive power setpoints constrained as $0 \le P_{i,t} \le \lambda_t P_i^{max}$ and $-Q_i^{max} \le Q_{i,t} \le Q_i^{max}$, with $\lambda_t \in [0,1]$ denoting the time-varying solar irradiance. Grid topology changes are modeled via a status vector $L = [l_1, \dots, l_p]$, where each switch $l_i \in \{0,1\}$ indicates open/closed status. The PV control objective is to maximize generation while minimizing voltage violations:
\begin{equation*}
\begin{aligned}
&\max_{P_0,\dots,P_T, Q_0,\dots,Q_T} \mathbb{E}_{L \sim \mathcal{P}} \Big[ \sum_{t=0}^{T} C^\top P_t - N_\text{vvn} \Big],\\
& \quad \quad \quad \text{s.t. } V_t = \text{PF}(P_t, Q_t, P_t^\text{load}, Q_t^\text{load}),
\end{aligned}
\end{equation*}
where $P^{\mathrm{load}}_{i,t}$ and $Q^{\mathrm{load}}_{i,t}$ denote the active and reactive load injections, respectively, and act as topology-independent exogenous inputs to the power flow model, $\text{PF}(\cdot)$ is the power flow solution, and $N_\text{vvn}$ counts nodes violating voltage limits. To address the intractability of this problem under unknown topology distributions, an adaptive in-context learning framework is employed. The system state is defined as topology-agnostic load injections $X_t = [P_t^\text{load}, Q_t^\text{load}]$, while the inputs $U_{i,t} = [{\alpha^i_{k,t}}, {\beta^i_{k,t}}]$ and outputs $Y_{i,t} = V_{i,t}$ are topology-dependent, with PV setpoints $P^i_{k,t} = \alpha^i_{k,t}\lambda_t P_k^{max}$ and $Q^i_{k,t} = \beta^i_{k,t} Q_k^{max}$. Unknown dynamics $X_{t+1} = f(X_t)$ and $Y_{i,t} = g_i(X_t, U_{i,t})$ are approximated via a two-stage model: a long short-term memory (LSTM) defined as $G_{\tilde{\theta}}$ for load forecasting, and a decoder-only LLM $H_\theta$ trained with in-context learning as a topology-adaptive surrogate for power flow. The LLM is fine-tuned on a small dataset $\mathcal{D}^\text{fine-tune}_i$ for any unseen topology, yielding predicted outputs $\hat{Y}_{i,t} = H_\theta(X_t, U_{i,t} \mid D^\text{fine-tune}_i)$. These models are embedded within a receding-horizon MPC:
\begin{equation*}
\begin{aligned}
\min_{U_{i,t},\dots,U_{i,t+T_h}} \sum_{j=0}^{T_h} \|Y_{i,t+j} - Y^\text{ref}\|_Q^2 + \|U_{i,t+j}\|_R^2,
\end{aligned}
\end{equation*}
subject to predicted state evolution $X_{i,t+j+1} = G_{\tilde{\theta}}(\cdot)$, linearized in-context voltage predictions $Y_{i,t+j} = \nabla H_\theta(\cdot) + Y_\text{eq}$, and input constraints $U_{min} \le U_{i,t+j} \le U_{max}$. This in-context MPC scheme leverages the in-context learning ability of the LLM to develop a surrogate model, which, after fine-tuning with a few samples, can predict the power flow solution for any topology. Combined with a load forecaster and an MPC control module, this adaptive framework enables real-time adaptation to unseen topologies using minimal data, achieving superior voltage regulation and reduced PV curtailment.

\paragraph{Experiment and Simulation Variations:}
Recently, LLMs have been applied to explore simulation variations in control systems by reasoning over models, parameters, and outcomes. By using multi-agent architectures and ICL, they can parameterize simulations, infer system invariants, and iteratively propose modifications with minimal human intervention~\cite{r30,EXp1,EXp2,EXp3,EXp4}. For instance, a multi-agent LLM framework in~\cite{EXp1} dynamically adjusts process-simulation parameters within a digital twin. Agents4PLC and Spec2Control use LLM agents to generate and verify PLC code in a closed-loop architecture~\cite{EXp2,EXp3}, supporting virtual validation of control logic. An LLM-Agent-Controller system coordinates specialized agents to simulate and analyze controllers~\cite{r30}. In large-scale power networks, Grid-Agent~\cite{EXp4} combines semantic reasoning with numerical solvers to simulate and validate control actions in real time. Together, these examples illustrate how LLMs can actively simulate and optimize complex control systems with minimal human intervention.

\begin{figure}
\centering
\includegraphics[width=\columnwidth]{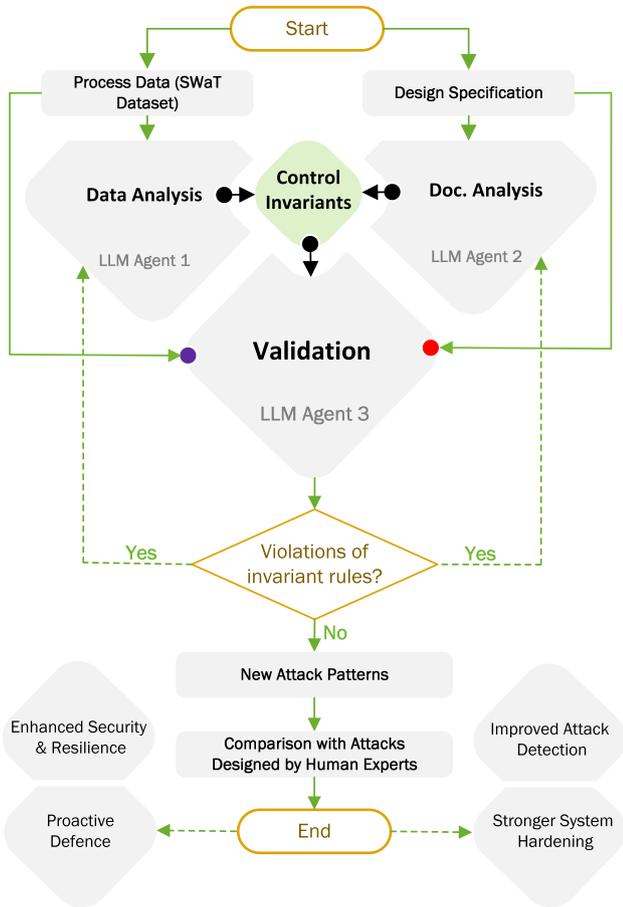}
\caption{LLM‑based workflow for generating and validating control invariants to derive new attack patterns.}
\label{fig9}
\end{figure}

Beyond standard simulation, LLMs can identify vulnerabilities and generate experimental or adversarial variations, offering exploratory flexibility beyond traditional control tools, particularly in safety-critical or resource-constrained systems~\cite{EXp5}. AttackLLM, for example, uses LLM agents to analyze operational datasets and design documentation, infer system invariants, and autonomously generate attack patterns. In the SWaT water treatment testbed, it produced 159 attack patterns compared to 36 manually designed by experts, capturing cross-stage dependencies and stealthy manipulations. The workflow in Fig.~\ref{fig9} illustrates how multiple LLM agents process datasets and design specifications, validate inferred invariants, and iteratively generate and refine attack patterns. By combining data- and design-driven techniques, AttackLLM produces diverse, high-quality sequences for assessing ML resilience in ICS, overcoming challenges such as limited testing environments, costly expert input, and sparse datasets. Overall, it demonstrates how LLMs can systematically explore simulation and experimental variations to uncover previously unknown vulnerabilities and deliver actionable insights for enhancing robustness.
\begin{figure}
\centering
\includegraphics[width=\columnwidth]{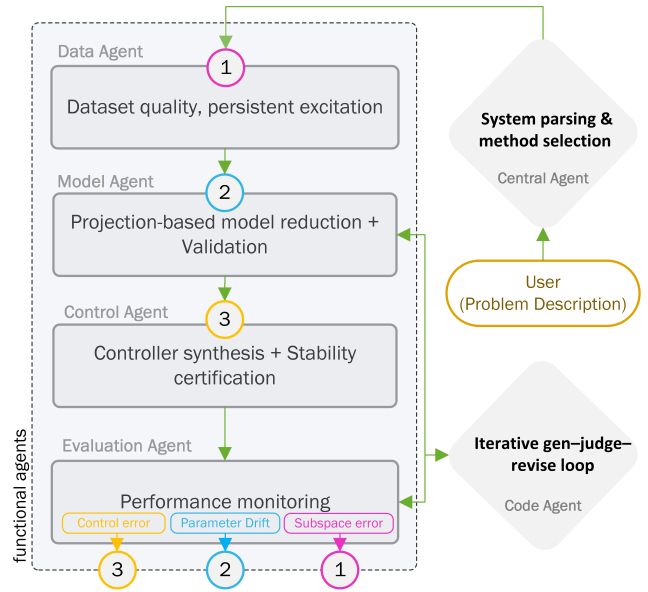}
\caption{An end-to-end LLM-based control architecture composed of two layers: functional agents and a code agent.}
\label{fig10}
\end{figure}

\subsubsection{Direct (An End-to-End LLM-Driven Control)}
Beyond the key functional roles, additional forms of LLM involvement in control are rapidly emerging, with recent research exploring whether several, or even all, of these functional categories can be integrated within a single unified framework~\cite{EndtoEnd1}. End-to-end LLM-driven control refers to architectures in which a multi-agent LLM system autonomously orchestrates the major stages of control design, including system analysis, modeling, controller synthesis, closed-loop evaluation, and iterative refinement. AURORA exemplifies this approach, employing specialized agents for system parsing, data generation, reduced-order model (ROM) construction, controller synthesis, and performance evaluation, while a shared Code Agent implements a generate–judge–revise cycle~\cite{EndtoEnd2}. The complete end-to-end workflow is depicted in the block diagram in Fig.~\ref{fig10}, which highlights how each agent interacts with explicit mathematical system representations, ROM projections, and reduced-order controller formulations such as MPC and adaptive control laws. Through sequential collaboration and continuous refinement, the framework achieves a high degree of autonomy in controller design, ensuring stable closed-loop performance while detecting and correcting degradation during deployment. Empirical results indicate the LLM-driven agents in this system achieve stable closed-loop control on benchmark problems, improving tracking performance by 6-12\% over expert-designed baselines.

\subsubsection{Direct (Domain-Specific Applications)}
As discussed above, LLMs for control span core functional roles and can directly influence controller design, demonstrating capabilities that go far beyond traditional research assistance. In such direct applications, LLMs act like junior control engineers, producing tuned parameters, validated models, and executable control policies. Conversely, in indirect applications, LLMs serve as research assistants that support concept exploration, system analysis, and literature-driven reasoning, thereby accelerating higher-level workflow development. These two modes are complementary and together characterize the emerging integration of LLMs into modern control engineering. While the preceding sections illustrated LLM functionalities through representative examples, related studies have explored these concepts primarily from survey perspectives across a wide range of application domains, including building energy management systems (EMS)~\cite{appg1}, power systems~\cite{appg2,appg22,appg222,appg2222}, robotics~\cite{appg3,appg33}, transportation~\cite{appg4}, industrial automation~\cite{DSEddc4}, biology~\cite{appg6}, cybersecurity~\cite{appg7}, aerospace~\cite{appg8,appg9}, marine systems~\cite{appg10}, and other emerging areas~\cite{DSEres12,appg11,appg12}. Table~\ref{tab:llm_control_applications} compiles representative studies from this literature, summarizing their control tasks, LLM functionalities, contributions, and limitations.

\begin{table*}
\caption{Representative domain-specific applications of LLMs in control.}\label{tab:llm_control_applications}
\resizebox{\textwidth}{!}{\selectfont
\begin{tabular*}{\tblwidth}{p{0.6cm} p{1.2cm} p{1.5cm} p{2cm} p{5.9cm} p{3.8cm}}
\toprule
\textbf{Ref.} & 
\textbf{Domain} & 
\textbf{LLM Role} & 
\textbf{Control Task} & 
\textbf{LLM + Control Contribution} & 
\textbf{Limitation} \\
\toprule
~\cite{app1} &
EMS &
Modeling \& Simulation & 
Autom. building energy model &
Converts natural-language building descriptions to EnergyPlus models with 100\% accuracy &
Limited to simple geometries; complex zoning not supported \\
\midrule
~\cite{app11111} &
Architect. & 
Design exploration &
Building massing optimization &
LLMs process site parameters \& simulation outputs to generate optimized architectural forms &
Limited dataset diversity; poor generalization \\
\midrule
~\cite{EXp4} &
Power Grid &
Feedback design & 
Violation detection \& remediation &
LLM-guided planning combined with control-based power flow, safety rollback, adaptive network encoding, and violation resolution &
Computational complexity; real-time constraints require validation \\
%NARRATE~\cite{app3} &
%Robotic manipulation &
%Symbolic reasoning &
%Complex manipulation tasks &
%Interfaces LLM with MPC to generate objectives \& constraints; supports human feedback; improves success rate &
%Safety not formally guaranteed; limited visual feedback; estimation errors \\
%MultiBotGPT~\cite{app4} &
%Multi-robot control &
%Task variation &
%Multi-robot task assignment &
%GPT-3.5 translates natural language commands into executable multi-robot tasks; coordinates UAV–UGV actions &
%Limited formal safety guarantees; depends on LLM interpretation accuracy \\
\midrule
~\cite{app5} &
Robotics &
Software Design &
Assembly and sorting tasks &
Automates control software development; generates assembly paths via prompt optimization &
Non-optimal; developer corrections; manual compilation \\
\midrule
~\cite{app6} &
Transport. &
Simulation \& Experim. &
Behavior modeling \& planning &
Uses LLM agents as behaviorally rich traveler proxies in agent-based models &
Behavioral alignment issue; needs extensive validation \\
\midrule
~\cite{app7} &
Biology &
Symbolic reasoning &
Cellular behavior steering &
Translates language prompts into spatial interventions; enables control of cellular collectives &
Multi-behavior learning interference; scalability issues \\
\midrule
~\cite{app8} &
IoT &
Feedback design &
Cyber-physical defense &
Fuses LLM semantic reasoning with multi-agent RL; enables autonomous, context-aware defense &
Requires high-quality data; sim-to-real \& robustness issue \\
\midrule
~\cite{app9} &
Aerospace &
Symbolic reasoning &
UAV mission execution &
Translates high-level commands into executable UAV mission code; integrates with autopilot &
Needs accurate code generation; limited perception \\
\midrule
%~\cite{app99} &
%Commercial aerospace &
%Design optimization &
%Spacecraft design and planning &
%LLMs analyze historical data to optimize spacecraft design and plan multi-objective missions &
%Conceptual demonstrations; practical deployment details needed \\
%CORALL~\cite{app10} & 
%Maritime ASVs &
%Decision exploration &
%COLREGs-compliant navigation &
%High-level LLM decision-maker with low-level planning; human-interpretable reasoning for maneuvers &
%Relies on general reasoning; fuzzy logic may miss environmental factors \\
%AIS-LLM~\cite{app10} &
%Maritime traffic & 
%Data synthesis &
%Trajectory prediction &
%Unified LLM framework for vessel prediction and anomaly detection; provides natural language explanations &
%Focused on simulation; real-world deployment requires adaptation \\
~\cite{app11} &
Maritime & 
Symbolic reasoning &
Route optimization &
Interprets language rules \& domain knowledge, collision avoidance, \& regulatory compliance &
Evaluated on benchmarks; no real-time integration\\
\bottomrule
\end{tabular*}}
\end{table*}

Among these applications, robotics is the most mature domain of LLM-enabled control, warranting dedicated discussion~\cite{RS1,RS2,RS3}. In this field, research has shifted over the past two years from employing LLMs as high-level assistants to integrating them directly into the action-generation loop, enabling real-time decision-making, control-policy synthesis, and task execution. LLMs now interpret tasks, generate and manage the control stack, produce motor primitives, sequence behaviors, evaluate execution, and perform closed-loop refinement. Reflecting these advances, contemporary robotic literature can be organized into several functional categories.
\begin{enumerate}[\textbullet] % from https://github.com/GT-RIPL/Awesome-LLM-Robotics
    \item \textbf{LLM-based reasoning:} Models that perform spatial, temporal, and semantic reasoning, infer object affordances, and provide contextual grounding to support planning and task execution~\cite{R-RS1,R-RS2,R-RS3,R-RS4,R-RS5,R-RS6,R-RS7,R-RS8,R-RS9,R-RS10,R-RS11}.
    \item \textbf{Task planning:} Systems that generate high-level symbolic or motion plans, including task-and-motion planning pipelines for complex sequences of actions~\cite{P-RS1,P-RS2,P-RS3,P-RS4,P-RS5,P-RS6,P-RS7,P-RS8,P-RS9,P-RS10,P-RS11,P-RS12,P-RS13,P-RS14,P-RS15,P-RS155}.
    \item \textbf{Manipulation:} Models that integrate LLMs with perception and affordance representations to perform grasping, placing, tool usage, and other complex manipulative skills~\cite{M-RS1,M-RS3,M-RS4,M-RS5,M-RS6,M-RS7,M-RS8,M-RS9,M-RS10,M-RS11}.
    \item \textbf{Navigation:} Models that interpret language or visual cues to guide autonomous navigation, locomotion, or human-robot interaction~\cite{R-RS2,N-RS2,N-RS3,N-RS4}.
    \item \textbf{Simulation frameworks and testbeds:} Platforms that enable LLM-driven experimentation, benchmarking, and RL in realistic or large-scale simulated areas~\cite{S-RS1}.
    \item \textbf{Safety, risk mitigation, and adversarial testing:} Research addressing robustness, bias, formal guarantees, and red-teaming of LLM-driven robotic systems~\cite{SAfty-RS1,SAfty-RS2,SAfty-RS3,SAfty-RS4,SAfty-RS5,SAfty-RS6,SAfty-RS7}.
\end{enumerate}
This taxonomy highlights how robotics has become the leading testbed for embodied LLM control, offering numerous concrete implementations in direct policy generation, manipulation, navigation, multimodal grounding, and online corrective reasoning. Compared with other application domains, robotic systems provide the clearest evidence that LLMs can act not only as high-level reasoning engines but as components embedded inside the control loop itself.

\subsection{Control for LLM}
Despite their strong generative abilities~\cite{CtoLLM0}, ensuring post-deployment accuracy, safety, and compliance remains challenging for LLMs. Traditional benchmarks such as HellaSwag~\cite{CtoLLM1}, MMLU~\cite{CtoLLM2}, TruthfulQA~\cite{CtoLLM3}, and MATH~\cite{CtoLLM4} assess static performance on isolated question-answer pairs but fail to capture the dynamic, interactive behavior of real-world LLM systems. In practice, users engage in multi-turn interactions, iteratively refining instructions and correcting deviations. With large context windows, LLMs accumulate internal state across turns, making their behavior sequential and state-dependent. As demand grows, attribute-specific, constraint-aware, and controllable text generations become essential for aligning outputs with desired properties~\cite{CtoLLM5,CtoLLM18}. As illustrated in Fig.~\ref{fig11}(a), LLMs can be modeled as dynamical systems that map an initial state $x_0$ to an output $y$ under control inputs $u$, representing user instructions or token-level interventions~\cite{Intro27}. From a control-theoretic perspective, this corresponds to a discrete-time stochastic system in which the state evolves token by token via the internal function $f_\theta$. Here, $f_\theta$ acts as the plant, with parameters $\theta$ defining the intrinsic dynamics and control inputs $u$ steering the system toward a desired target state, raising fundamental questions of controllability, cost, and computational complexity.

A wide range of methods has been proposed for controllable text generation, most of which rely on heuristic guidance rather than formal guarantees. Parameter-level approaches modify model weights to induce specific behaviors~\cite{CtoLLM9,CtoLLM10}, while activation-level steering intervenes at inference time by biasing internal representations toward desired regions~\cite{CtoLLM11,CtoLLM12,CtoLLM13}. In contrast, prompt engineering shapes outputs through carefully designed instructions without altering model parameters or activations~\cite{Intro27,CtoLLM7,CtoLLM8}. The following sections examine these three strategies in detail:
\begin{figure}
\centering
\includegraphics[width=\columnwidth]{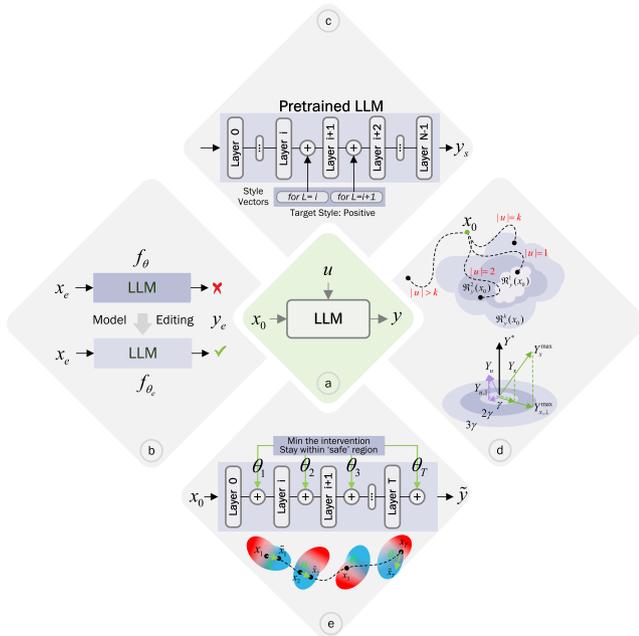}
\caption{Control theoretic perspective of LLMs.}
\label{fig11}
\end{figure}

\subsubsection{Model Editing} As seen from Fig.~\ref{fig11}(b), model editing can be interpreted as a controllable intervention in the parameter space, where changes to $\theta$ steer the model toward desired outputs while preserving unrelated behaviors. Formally, an edit introduces a change $\Delta \theta$ such that the edited model is
\begin{equation*}
\begin{aligned}
\theta \to \theta_e = \theta + \Delta \theta, \quad f_{\theta_e} = f_\theta + \Delta f_\theta.
\end{aligned}
\end{equation*}
The goal is to adjust the model $f_\theta : \mathcal{X} \to \mathcal{Y}$ on a specific edit descriptor $(x_e, y_e)$, producing a post-edit model $f_{\theta_e}$ such that $f_{\theta_e}(x_e) = y_e$ while maintaining its behavior on unrelated inputs:
\begin{equation*}
    f_{\theta_e}(x) =\begin{cases} y_e, \quad if \, \,  x \in i(x_e, y_e) \\
    f_\theta(x), \quad if \, \,   x \in o(x_e, y_e),
    \end{cases}
\end{equation*}
where $i(x_e, y_e)$ denotes the in-scope region, typically including $x_e$ and semantically related inputs $n(x_e, y_e)$, and $o(x_e, y_e)$ denotes the out-of-scope region of unrelated inputs. A successful model edit is typically evaluated along three axes~\cite{CtoLLM9}: reliability, generalization, and locality. Reliability measures whether the edit produces the correct output for the original example $(x_e, y_e)$. Generalization evaluates whether the edit extends to semantically related inputs in the equivalence neighborhood $n(x_e, y_e)$. Locality ensures that predictions on out-of-scope inputs remain unchanged, preserving the model's behavior on unrelated data. Model editing methods for LLMs can be broadly classified into parameter-preserving and parameter-modifying approaches:
\begin{enumerate}[\textbullet]
\item Parameter-preserving methods store edit information externally and guide the model without changing its parameters $\theta$. This technique includes memory-based approaches, which retrieve stored edits to influence outputs, and extra-parameter approaches, which add trainable modules or codebooks activated for specific edits.
\item Parameter-modifying methods directly update $\theta$ to enforce edits. Locate-and-edit techniques identify neurons or matrix entries to modify, and meta-learning or hypernetwork approaches predict low-rank updates to achieve targeted edits while preserving unrelated behavior.
\end{enumerate}
While enabling precise and controllable text generation, model editing is complex and carries risks such as unintended side effects and architectural instability, with trade-offs in reliability, generalization, and locality.

\subsubsection{Activation Engineering} This approach provides state-feedback control for LLMs. Instead of modifying parameters, a style-specific steering signal $v_s^{(i)}$ is injected into hidden activations $a^{(i)}(x)$ at layer $i$ during the forward pass (see Fig.~\ref{fig11}(c)):
\begin{equation*}
\begin{aligned}
\hat{a}^{(i)}(t) = a^{(i)}(t) + \lambda v_s^{(i)},    
\end{aligned}
\end{equation*}
where $\lambda$ controls the influence of the style vector, enabling nuanced steering toward a desired style category $s \in S$, such as positive/negative sentiment or emotions~\cite{CtoLLM11}. The steering vector $v_s^{(i)}$ can be obtained using two main approaches:
\begin{enumerate}[\textbullet]
\item Training-based style vectors: A vector is optimized so that, when injected at layer $i$, the model generates sentences of a target style. Sentence-specific vectors are aggregated across examples within a style category $s$, and contrasted with vectors from other styles to produce the final steering vector. This approach does not require knowing the original style of the input.
\item Activation-based style vectors: Vectors can also be derived directly from the model's hidden activations. Layer activations are averaged over sentences in style $s$ and contrasted with activations from other styles to produce the steering vector. This method generalizes prior pairwise contrast approaches and eliminates the need for optimization.
\end{enumerate}
Activation engineering enables real-time, continuous adjustment of the model style, with the scaling parameter $\lambda$ controlling the strength of the steering effect. Compared to model editing, it provides finer-grained and reversible control without modifying parameters. It allows precise and transparent manipulation of internal representations, though its effects are temporary and can degrade semantic coherence if over-steered. It currently lacks a formal theory of reachability, with effectiveness dependent on local representation geometry.

\subsubsection{Prompt Engineering (Input Optimization)} Input optimization can be viewed as an open-loop feedforward control mechanism, where an external sequence of control tokens $u(0),\cdots,u(k)$ initializes and shapes the state trajectory~\cite{Intro27}. A causal LLM can be denoted as $P_{LM}$, which maps an ordered list of tokens from a vocabulary $\mathcal{V}$ to a probability distribution over the next token. For a sequence $\mathbf{x} = [x^1, \dots, x^n] \in \mathcal{V}^n$, where $x \in \mathcal{V}$, the model defines
\begin{equation*}
P_{LM}(x_{n+1} \mid \mathbf{x}) \in [0,1]^{|\mathcal{V}|}.
\end{equation*}
Although LLMs are sometimes modeled as generating outputs in a single step, token generation and externally imposed control inputs actually occur sequentially, producing non-trivial system dynamics. At each time step $t$, the newest token is either sampled from a control input $u(t)$ or generated by the model, i.e., $x' \sim P_{LM}(x' \mid \mathbf{x}(t))$. Unlike classical ordinary differential equations, LLMs operate on discrete states and discrete time, and their state $\mathbf{x}(t)$ grows as tokens are added rather than remaining fixed. This gives rise to a shift-and-grow dynamic, in which each token is determined either by control or by the model's intrinsic generation. An autoregressive LLM system with control input can be defined as $\Sigma = (\mathcal{V}, P_{LM})$ over the time set $T = \mathbb{N}$, with state space $\mathcal{X} = \mathcal{V}^*$, the set of all possible token sequences. The state at time $t$ is $\mathbf{x}(t) = [x^0(t), \dots, x^t(t)]$, and the input space is $U = \mathcal{V} \cup \varnothing$. The base model transition map $\phi$ updates the state as
\begin{equation*}
\phi(\mathbf{x}(t), u(t), t, t+1) =
\begin{cases}
\mathbf{x}(t) + u(t), \quad \text{if} \, \, \, u(t) \neq \varnothing \\
\mathbf{x}(t) + x', \quad \, \, \, \, \, \text{otherwise}.
\end{cases}
\end{equation*}
Multi-step transitions $\phi(\mathbf{x}(t), u(t), t, t+N)$ are obtained by iterating this map for control sequences $u$ defined over the interval $[t, t+N]$. A readout map $h$ returns the last $r$ tokens of a state sequence as $h(\mathbf{x}(t); r) = [x^{t-r}(t), \dots, x^t(t)]$.

This framework generalizes naturally to LLM augmentations such as CoT, retrieval-augmented generation, and chatbot interactions. For instance, CoT corresponds to sampling the readout map $h(\mathbf{x}(T), r)$ at $T > |\mathbf{u}| + |\mathbf{x}_0| + r$ for a prompt $\mathbf{u}$ and initial state $\mathbf{x}_0$. At each step, the control input $\mathbf{u}(t)$ either allows token generation $\mathbf{u}(t) = \varnothing$ or overrides it with a token $\mathbf{u}(t) \neq \varnothing$. Finite-length inputs $\mathbf{u} \in \mathcal{V}^k$ are implicitly zero-padded. Although next-token generation $x'$ is probabilistic, assuming greedy decoding allows analysis of \textit{reachable sets} and \textit{controllability} (see Fig.~\ref{fig11}(d)):
\begin{enumerate}[\textbullet]
\item The reachable output set for LLM system $\Sigma = (\mathcal{V}, P_{LM})$, denoted by $R_y(\mathbf{x}_0)$, consists of all reachable outputs $y \in \mathcal{V}^*$ from $\mathbf{x}_0$.
\item This system is output-controllable if $R_y(\mathbf{x}_0) = \mathcal{V}^*$ for every $\mathbf{x}_0$.
\item Under prompt-length constraints $|\mathbf{u}| \le k$, the system $\Sigma = (\mathcal{V}, P_{LM})$ is $k-\epsilon$ controllable with respect to a dataset $\mathcal{D} = \{(\mathbf{x}_0^i, \mathbf{y}^i)\}$, for $i \in [N]$, if $\Pr[\mathbf{y} \notin R_y^k(\mathbf{x}_0)] \le \epsilon$, for $(\mathbf{x}_0, \mathbf{y}) \sim \mathcal{D}$.
\end{enumerate}
Empirical analysis of $k-\epsilon$ controllability is challenging due to the large, unstructured state space. Optimal prompts $\mathbf{u}^*$ can provide a lower bound on the reachable set, while theoretical results provide upper bounds for self-attention. For a self-attention layer, the mapping $\Xi:\mathbb{R}^{N \times d_{in}} \to \mathbb{R}^{N \times d_{out}} $ is
\begin{equation*}
\Xi(X;\theta) = D^{-1} \exp(\frac{QK^\top}{\sqrt{d_\text{key}}}) V,
\end{equation*}
where $Q = XW_q$, $K = XW_\text{key}$, and $V = XW_v$ with weight matrices $\theta = (W_q, W_\text{key}, W_v)$ and
\begin{equation*}
D := \text{diag}(\exp(\frac{QK^\top}{\sqrt{d_\text{key}}}) 1_{N \times 1}),
\end{equation*}
where $1_{N \times 1}$ is an $N \times 1$ matrix of ones. For a partitioned input $X = [U; X_0]$ with $k$ control tokens $U$ and $m$ imposed tokens $X_0$, the output is $X' = [U'; Y]$, where $Y \in \mathbb{R}^{m \times d_\text{out}}$ represents the target output. Decomposing $Y$ as
\begin{equation*}
\begin{aligned}
    Y &= Y_u + Y_x 
      &= (Y_{u,\parallel} + Y_{u,\perp}) + (Y_{x,\parallel} + Y_{x,\perp}),
\end{aligned}
\end{equation*}
the \textit{self-attention control theorem} states that the desired output $Y^*$ is unreachable for any control $U$ if $\|Y^{\max}_{x,i,\perp}\| > k \gamma_i(X_0, \theta), \quad i \in {1,\dots,m}$~\cite{Intro27}, where
\begin{equation*}
\begin{aligned}
\gamma_i(X_0, \theta) &:= \frac{e^\alpha}{g_i} \sigma_v M_u, \quad  \alpha = \frac{\sigma_q \sigma_\text{key} M_u M_x}{\sqrt{d_\text{key}}}, \\
g_i(X_0, \theta) &:= \sum_{j=1}^m \exp\Big(\frac{(X_0)^i W_q W_{\text{key}}^\top (X_0)^{j\top}}{\sqrt{d_\text{key}}} \Big),
\end{aligned}
\end{equation*}
and $\sigma_v$, $\sigma_q$, and $\sigma_\text{key}$ are the maximum singular values of the value, query, and key projections, respectively, with $M_u := \max_j |U^j|$ and $M_x := \max_j |(X_0)^j|$. This bound scales linearly with $k$, so increasing the number of control tokens reduces the set of unreachable outputs $Y^*$, which highlights that longer control inputs expand the effective reach of self-attention layers.

\begin{table*}
\caption{Control-theoretic comparison of model editing (ME), activation engineering (AE), prompt engineering (PE), and LiSeCo.}\label{tab:control_comparison}
\resizebox{\textwidth}{!}{\selectfont
\begin{tabular*}{\tblwidth}{p{0.8cm} p{1.9cm} p{2.2cm} p{1.2cm} p{2.6cm} p{2.9cm} p{3.1cm}}
\toprule
\textbf{Method} &
\textbf{Control type} &
\textbf{Actuation Point} &
\textbf{Longevity} &
\textbf{Objective} &
\textbf{Strengths} &
\textbf{Limitations} \\
\toprule
\textbf{ME} &
Parameter-space&
Parameter update by $\Delta \theta$ &
Permanent &
Function correction with locality preserv. &
Strong locality; Precise behavioral correction &
Limited scale and editing scope; Ethical limits \\
\midrule
\textbf{AE} &
State-feedback &
Activ. perturb. by $\lambda v_s^{(i)}$ &
Temporary &
Trajectory steering to style manifolds &
High style controllability; Smooth modulation; Enables new styles &
High cost; Sentiment/emotion focused; Complex/multidim. styles\\
\midrule
\textbf{PE} &
feedforward &
External prompt sequence $u(t)$ &
Temporary &
Reachability via prompt design &
Strong token controllability; Useful prompt optimization insights &
Limited multi-token control; Model-specific generality; Partial reach. \\
\midrule
\textbf{LiSeCo} &
State-feedback (optimal cont.) &
Activ. intervention $x_t^\text{ctrl} = x_t + \theta_t^*$ &
Temporary &
Trajectory steering into safe semantic regions &
Provable guarantees; Fine-grained, context-dependent control &
Offline training required; Limited to pre-defined attribute regions\\
\bottomrule
\end{tabular*}}
\end{table*}

While steering-based methods work well in practice, they generally lack formal guarantees of success~\cite{Intro2333}. Most approaches operate by nudging latent representations along learned directions without ensuring that outputs satisfy desired constraints, and their overall effectiveness remains poorly understood~\cite{CtoLLM144}. Editing-based methods can introduce unintended global effects, while prompt-based interventions (e.g., ReFT) act only on initial representations and cannot influence activations during generation~\cite{CtoLLM13,CtoLLM15}. In short, true control with enforceable hard constraints and verifiable guarantees remains largely unrealized. Recent studies have framed LLMs as control systems: BWArea models generation via world and inverse dynamics with a guiding policy~\cite{CtoLLM17}, control barrier functions constrain decoding to safe outputs~\cite{CtoLLM19,CtoLLM20}, and representation editing computes optimal latent-space interventions in discrete-time stochastic systems~\cite{CtoLLM21}. In another influential study, the authors in~\cite{Intro2333} introduce linear semantic control (LiSeCo), framing online LLM intervention as an optimal control problem over activations. Treating the LLM as a dynamical system $x_{t+1} = f_\theta(x_t)$, LiSeCo computes minimal, provably correct latent-space interventions that steer trajectories into predefined safe semantic regions $\mathcal{X}_t \subset \mathbb{R}^d$ without altering $f_\theta$ (see Fig.~\ref{fig11}(e)). LiSeCo operates in two phases, analogous to a sensor–controller architecture:
\begin{enumerate}[\textbullet]
\item Offline Phase (Probing the Safe Region): For each layer $t$, a linear probe $g_t(x_t) = \nu(W_t^\top x_t)$ is trained to map activations to a target attribute score. The safe region is then defined as $\mathcal{X}_t = \{ x \in \mathbb{R}^d \mid \alpha^{\min} \leq \nu(W_t^\top x) \leq \alpha^{\max} \}$, where $[\alpha^{\min}, \alpha^{\max}]$ is the desired attribute range.
\item Online Phase (Closed-Form Activation Control): During generation, at each forward pass, the current activation $x_t$ is checked. If it lies outside $\mathcal{X}_t$, LiSeCo solves a constraint optimization problem and computes an optimal control input $\theta_t^*$ and intervenes in the latent space $x_t^\text{ctrl} = x_t + \theta_t^*$. The model then propagates this controlled activation via its unchanged function $x_{t+1} = f_\theta(x_t^\text{ctrl})$.
\end{enumerate}
This closed-form intervention is computationally efficient and applied per forward pass, preserving the quality and naturalness of generated text. In comparison, prompting acts on the input $u(t)$, leaving $f_\theta$ unchanged and providing only temporary control; activation steering heuristically perturbs hidden activations, also without modifying $f_\theta$ and with ephemeral effects; and model editing directly alters parameters $\theta$, permanently changing $f_\theta$. LiSeCo, in contrast, optimally intervenes on activations $x_t$ with provable guarantees, leaving $f_\theta$ intact while temporarily steering trajectories into desired semantic regions. In this sense, LiSeCo functions as optimal activation steering, offering fine-grained, context-dependent control while balancing reliability, expressivity, and computational efficiency (see Table~\ref{tab:control_comparison} for a quick overview).

\subsection{Control with LLM}
A control-theoretic perspective provides a principled framework for analyzing the dynamics of LLMs. In this view, hidden activations across layers constitute the system state, input tokens act as control inputs, and output token probabilities represent the system output. To date, research at the intersection of control and LLMs has primarily focused on two directions: using LLMs for control, and applying control to LLMs to shape their outputs. In this section, we adopt a control-theoretic perspective to analyze LLMs themselves. By bridging sequence modeling and classical control, we aim to uncover structural and dynamical insights into intrinsic model behavior, focusing on the integration of control principles within LLMs. Over the past two decades, deep learning has driven significant advances across diverse fields, underscoring the importance of structured model architectures~\cite{ContwithLLM1}. Recurrent NNs (RNNs) have been widely used for sequential tasks such as machine translation, but they struggle with long-range dependencies, sequential computation bottlenecks, and capturing global context due to the lack of attention mechanisms~\cite{ContwithLLM2}. RNNs process sequences by maintaining a hidden state that acts as memory. At each time step $k$, given input $x_k \in \mathbb{R}^D$ and previous hidden state $h_{k-1} \in \mathbb{R}^N$, the hidden state and output $o_k \in \mathbb{R}^O$ are updated as
\begin{equation*}
\begin{aligned}
    h_k &= \tanh(W_{hx} x_k + W_{hh} h_{k-1} + b_h),\\
    o_k &= W_{oh} h_k + b_o,
\end{aligned}
\end{equation*}
where $W_{hx} \in \mathbb{R}^{N \times D}$, $W_{hh} \in \mathbb{R}^{N \times N}$, $W_{oh} \in \mathbb{R}^{O \times N}$, and $b_h, b_o$ are biases. While effective for sequential data, RNNs face challenges with long-range dependencies, sequential computation, and lack of attention. Transformer-based models address these limitations by using self-attention to focus on relevant parts of the input and capture long-range dependencies, forming the foundation of LLMs like ChatGPT~\cite{ContwithLLM3}. As discussed earlier, given queries $Q$, keys $K$, and values $V$, attention scores are computed via the scaled dot-product and softmax, $S = \text{Softmax}\Big(\frac{Q K^\top}{\sqrt{d_K}}\Big)V$, and multi-head attention aggregates multiple heads as $(S_1 \oplus S_2 \oplus \dots \oplus S_m)W_o$, where $\oplus$ denotes concatenation and $W_o$ is a projection matrix. Although Transformers effectively capture global dependencies, their self-attention mechanism incurs a quadratic computational cost, limiting efficient long-context processing~\cite{ContwithLLM4}. Despite their success, self-attention offers limited controllability: recent studies~\cite{Intro27} model LLMs as discrete stochastic dynamical systems and analyze controllability through attention, but interventions currently provide steering rather than guaranteed control. Key open questions in LLM control include:
\begin{enumerate}[\textbullet]
\item CoT Control: Analyzing stability, reachability, and composability when LLMs generate intermediate tokens.
\item Distributional Control: Steering the output distribution $P_{\mathrm{LM}}(y \mid x_0 + u)$ toward a desired target $P^*(y)$.
\item Computational Cost: Trade-offs between control performance and efficiency.
\item Learnability of Control: How effectively can LLMs learn to control other LLMs?
\item Controllable Subspaces: Identifying controllable versus uncontrollable components in Transformers.
\item Composable Systems: Designing modular, predictable LLM control architectures.
\end{enumerate}

Recent work such as LiSeCo explores activation-level control and provides theoretical guarantees~\cite{Intro2333}, but its effectiveness is limited by the need for supervised probes and representative training data. Structured state space models (SSMs)~\cite{ContwithLLM5} have recently emerged as efficient alternatives to Transformers, capturing complex dependencies via linear, well-characterized dynamics inspired by classical SSMs~\cite{ContwithLLM6}. Combining recurrent and convolutional architectures, SSMs achieve linear or near-linear scaling with sequence length, reducing computational cost. The continuous-time SSM is defined as
\begin{equation*}
\begin{aligned}
\dot{h}(t) &= A h(t) + B x(t),\\
y(t) &= C h(t) + D x(t),
\end{aligned}
\end{equation*}
where $x \in \mathbb{R}$ and $y \in \mathbb{R}$ are the input and output, $h \in \mathbb{R}^N$ is the hidden state, $A \in \mathbb{R}^{N \times N}$ is the state transition matrix, $B \in \mathbb{R}^{N \times 1}$ is the input matrix, $C \in \mathbb{R}^{1 \times N}$ is the output matrix, and $D \in \mathbb{R}$ controls direct input-to-output effects. Typically, $Dx(t)$ is omitted, acting as a skip connection. After discretization using zero-order hold with time step $\Delta$, one has $h_k = \Bar{A} h_{k-1} + \Bar{B} x_k$ and $y_k= C h_k$, with $\Bar{A} = exp(\Delta A)$ and $\Bar{B} = (\Delta A)^{-1}(\Bar{A} -I)\Delta B$. This structure allows both recurrent and convolutional computation as $y = x \Bar{K}$, $\Bar{K} = \{C\Bar{B}, C\Bar{A} \Bar{B}, \dots, C\Bar{A}^k \Bar{B},\dots\}$, enabling parallel GPU computation and linear scalability with sequence length.

\begin{figure}
\centering
\includegraphics[width=\columnwidth]{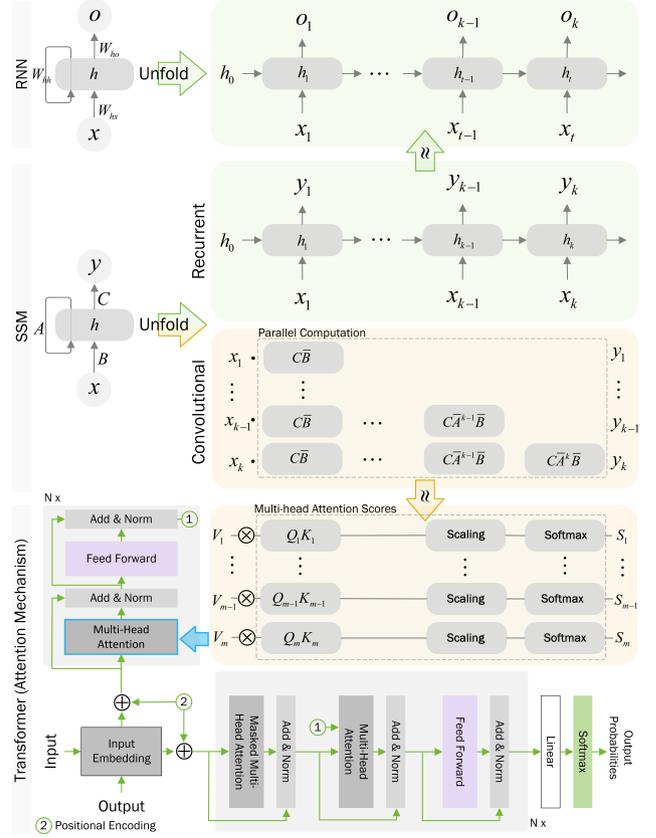}
\caption{Representative model architectures: RNNs, Transformers, and SSMs.}
\label{fig12}
\end{figure}

Figure~\ref{fig12} compares the computation patterns of RNNs, Transformers, and SSMs. RNNs rely on nonlinear recurrent updates, enabling fast autoregressive outputs but with limited parallelism and slower training. Transformers compute large matrix multiplications across query–key pairs in parallel, allowing efficient training but slow autoregressive inference. Discrete SSMs can operate in either recurrent or convolutional form due to their linear structure, supporting both efficient inference and parallel training. However, conventional time-invariant SSMs with input-independent parameters $(A, B, C, \Delta)$ are limited in modeling context, reducing performance on tasks such as selective copying~\cite{Intro25}. The central challenge in sequence modeling is compressing context into a smaller state. Transformers capture long-range dependencies effectively but store the entire sequence during inference, making them inefficient. Recurrent models and SSMs like S4 are more efficient but limited in context compression. Tasks such as selective copying and induction Heads highlight these trade-offs, as LTI models cannot implement input-dependent dynamics and static convolutions struggle with variable input–output spacing~\cite{ContwithLLM1}. Mamba addressed these limitations with selective SSMs featuring input-adaptive parameters and a hardware-aware recurrent scan, achieving strong long-range performance~\cite{Intro25}. Building on this, Mamba2, inspired by linear attention, demonstrates that SSMs can now achieve Transformer-level performance~\cite{ContwithLLM9}.

\subsubsection{Mamba}
To empower SSMs with Transformer-like capabilities, three key innovations are introduced in Mamba~\cite{Intro25}: High-order polynomial projection pperator (HiPPO)-based memory initialization, a selection mechanism, and hardware-aware computation (see Fig.~\ref{fig13}). Mamba implements selective SSMs (S-SSMs), where state-space parameters dynamically adapt to the input. The core update equations are
\begin{equation*}
\begin{aligned}
h_t &= A h_{t-1} + B x_t,\\
y_t &= C h_t + D x_t,
\end{aligned}
\end{equation*}
where all variables and parameters are defined as in the preceding section. Mamba focuses primarily on initializing the hidden-state matrix $A$ to capture rich temporal dependencies. This is achieved using HiPPO theory with a scaled Legendre measure (LegS), which enables the model to encode the entire historical context rather than relying on a limited sliding window. Building on HiPPO theory, Mamba adopts two simple initialization schemes for the hidden-state matrix $A$, where the $n$-th element of $A$ is defined as $-\frac{1}{2} + n i$ for the complex case and $-(n + 1)$ for the real case.

These initializations enable the model to encode long-range dependencies by applying slower decay to recent inputs and faster decay to older ones, effectively compressing and reconstructing historical information. HiPPO-LegS further provides favorable theoretical properties, time-scale consistency, fast computation, and bounded gradients and approximation error, making parameter learning more stable and efficient. Matrices $B$ and $C$ are rendered input-selective via $(B,C) \rightarrow S^{(B,C)} = W^{(B,C)} x$, and $\Delta \rightarrow S^\Delta = \tau_\Delta \cdot \text{BroadCast}_D(W^\Delta x)$, where $S^{(B,C)} \in \mathbb{R}^{M \times L \times N}$, and $S^{\Delta} \in \mathbb{R}^{M \times L \times D}$ dare input-dependent selective matrices that enable content-aware modeling. Here, $M$, $L$, $D$, and $N$denote the batch size, sequence length, input dimension, and hidden dimension. The matrices $W^B \in \mathbb{R}^{N \times D}$, $W^C \in \mathbb{R}^{N \times D}$, and $W^{\Delta} \in \mathbb{R}^{D}$ are linear projection weights, and $\mathrm{BroadCast}_D$ denotes broadcasting along all $d = 1,\dots,D$ dimensions. The S-SSM is then discretized as
\begin{equation*}
\begin{aligned}
\Bar{A}& \;\rightarrow\; S^{\Bar{A}} = \exp(S^{\Delta} A),\\
\Bar{B}& \;\rightarrow\; S^{\Bar{B}} = (S_{\Delta} A)^{-1} \big(\exp(S^{\Delta} A) - I \big)S^{\Delta} S^B,
\end{aligned}
\end{equation*}
where $S^{\Bar{A}} \in \mathbb{R}^{M \times L \times D \times N}$ and $S^{\Bar{B}} \in \mathbb{R}^{M \times L \times D \times N}$ are the selective state-transition and input matrices, now explicit functions of the input $x$. Consequently, the discrete SSM becomes a linear time-varying (LTV) (i.e., content-aware) system $y = \mathrm{SSM}(A,B,C)(x)$ producing an output $y \in \mathbb{R}^{M \times L \times D}$ conditioned on the input sequence. Mamba's time-varying selection mechanism is analogous to attention in Transformers, as both use input-dependent projections to modulate interactions. This enables flexible, content-aware modeling at the cost of strict convolutional equivalence and some computational efficiency. From a control-theoretic perspective, these mechanisms convert an LTI structured SSM into a controlled, input-dependent LTV system and govern:
\begin{enumerate}[\textbullet]
\item Controllability: How inputs affect the hidden state?
\item Observability: How this state influences the output?
\item Stability: How memory decays or resets?
\item Gain scheduling: How the effective timestep adjusts?
\end{enumerate}

\begin{figure}
\centering
\includegraphics[width=\columnwidth]{Fig13_SSMr.png}
\caption{Overview of S-SSM with hardware-aware state expansions.}
\label{fig13}
\end{figure}

\subsubsection{Mamba-2}
Transformers have driven major advances in deep learning and inspired techniques such as parameter-efficient fine-tuning, catastrophic forgetting mitigation, and model quantization. To enable SSMs to benefit from these Transformer-derived tools, Mamba-2 introduces structured state-space duality (SSD), a unified framework that establishes a theoretical connection between SSMs and attention mechanisms~\cite{ContwithLLM10}. Formally, $y = \mathrm{SSD}(A,B,C)(x) = Px$, where $P$ is the sequentially semi-separable matrix representation of an SSM, with entries $P_{ji} = C_j^{\top} A_{j:i} B_i$. Here, $B_i$ and $C_j$ denote the selective state-space matrices associated with input tokens $x_i$ and $x_j$, respectively, and $A_{j:i}$ represents the selective hidden-state transition from position $i$ to $j$. SSD shows that both Transformer attention and LTV SSMs can be interpreted as semi-separable matrix transformations, and S-SSMs are equivalent to structured linear attention with a semi-separable masking matrix~\cite{Intro25}. Building on SSD, Mamba-2 introduces a hardware-efficient block decomposition that formulates SSM computation as operations on semi-separable matrices, partitioning the dynamics into diagonal and off-diagonal blocks. This design enables faster training than the parallel associative scan used in Mamba while maintaining performance competitive with Transformers.

\subsubsection{Mamba Block Design, Scanning, and Data Adaptation}  
\begin{figure}
\centering
\includegraphics[width=\columnwidth]{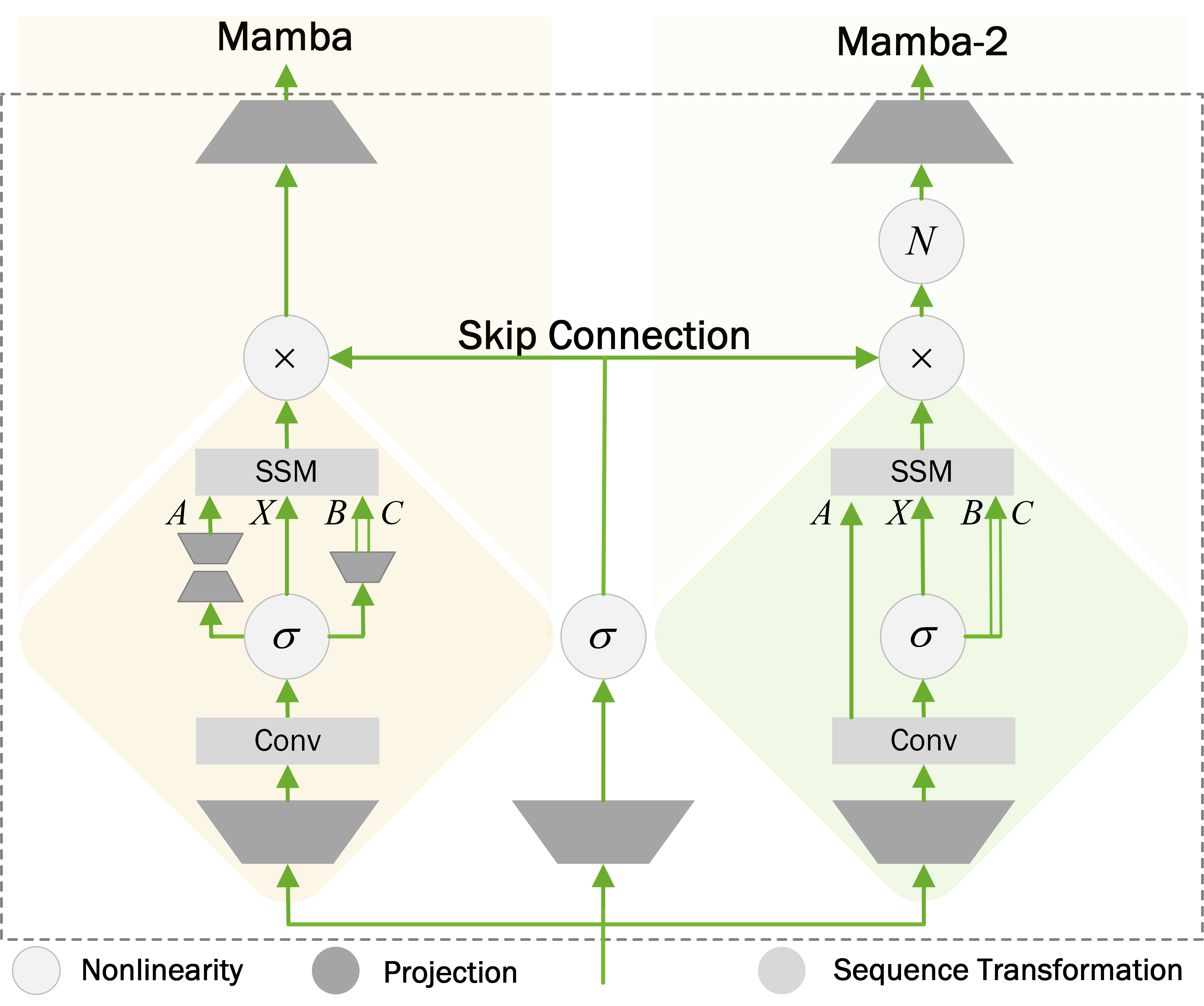}
\caption{Comparison of Mamba and Mamba-2 architectures.}
\label{fig14}
\end{figure}

The Mamba architecture, comprising Mamba and Mamba-2, employs S-SSM layers to efficiently map inputs to outputs. As shown in Fig.~\ref{fig14}, Mamba maps input sequences $X$ to outputs $Y$ via sequential linear projections through an S-SSM layer with a parallel associative scan, while Mamba-2 streamlines this by jointly projecting $[X, A, B, C]$ and adding post-skip normalization for improved stability and computational efficiency, enabling faster SSD computation than the parallel selective scan in Mamba. Enhancements to the Mamba architecture fall into three perspectives~\cite{ContwithLLM1}: block design, scanning modes, and memory management. Block design includes integration with Transformers, RNNs, and SNNs; use in frameworks such as U-Net and diffusion models; and techniques like mixture-of-experts to boost performance. Scanning modes improve input traversal and long-range dependency capture, while memory management ensures effective propagation of hidden states. Beyond sequential data, Mamba has been adapted to images, graphs, and 3D point clouds, demonstrating linear complexity and long-range modeling, and extended to multimodal learning for vision-language tasks, motion-language alignment, and fusion of complementary modalities. It has been demonstrated that Mamba achieves superior performance in both interpolation and challenging extrapolation tasks, consistently ranking among the top models while maintaining low computational cost and exceptional extrapolation capabilities~\cite{ContwithLLM100}.

\subsubsection{Control-Theoretic Properties of LLMs}
In classical control theory, controllable and observable canonical forms provide structured state-space representations that are fundamental to control system design. Building on this foundation, the \textit{sparse Mamba family} integrates control-theoretic structure directly into Mamba's state-space matrices~\cite{ContwithLLM11}, improving interpretability and enabling principled manipulation of system dynamics. When applied to Mamba LLMs, this approach enhances the controllability, observability, and stability of hidden states, establishing a more structured and analytically tractable framework for understanding LLMs behavior.
\paragraph{Controllability:} The controllable canonical form (CCF) is closely tied to reachability: a system is reachable if its state can be driven from any initial to any final condition in finite time, i.e., if the reachability matrix is full rank. CCF makes controllability explicit, simplifying controller design, state feedback, and pole-zero placement. For an LTI system with transfer function
\begin{equation*}
H(s)=
\frac{b_{n-1}s^{\,n-1}+b_{n-2}s^{\,n-2}+\cdots+b_1 s + b_0}
{s^n + a_{n-1}s^{\,n-1}+\cdots+a_1 s + a_0},
\label{eq:transfer}
\end{equation*}
the state matrix $A$ in the CCF is given as
\begin{equation*}
A_c=
\begin{bmatrix}
0 & 1 &  0 & \cdots & 0 & 0 \\
0 & 0 &  1 & \cdots & 0 & 0 \\
\vdots & \vdots & \ddots & \vdots & \vdots \\
0 & 0 & 0 & \cdots & 0 & 1 \\
-a_{n-1} & -a_{n-2} & -a_{n-3} & \cdots & -a_1 & -a_0
\end{bmatrix},
\label{eq:A_ccf}
\end{equation*}
with input matrix $B = [\, 0 \;\; 0 \;\; \cdots \;\; 1 \,]^{\top}$ and output matrix $C = [\, b_{n-1} \;\; \cdots \;\; b_0 \,]$. In sparse controllable-Mamba, a coefficient vector is initialized uniformly, embeded into the CCF structure during training. The matrix $B$ ensures full controllability, while $C$ determines how hidden states contribute to the output. The feed-through term $D$ is initialized to $0$ and treated as trainable. Any SSM may be converted into CCF via a similarity transformation.

\paragraph{Observability:} The observable canonical form (OCF) explicitly reveals a system's observability, i.e., whether its state can be determined from output measurements over a finite time interval, which holds if and only if the observability matrix has full rank. In OCF, the coefficients of the characteristic polynomial appear directly in the state matrix $A$ in the OCF is given as $A_o=A_c^{\top}$, with input matrix $B = [\, b_{n-1} \;\; \cdots \;\; b_0 \,]^{\top}$ and output matrix $C = [\, 0 \;\; \cdots \;\; 0 \;\; 1 \,]$. This form is the transpose of the CCF, reflecting the duality between controllability and observability. In sparse observable-Mamba, these structures are incorporated directly into the SSM parameters, enabling observer-like behavior. The scalar $D$ remains trainable.

\paragraph{Stability:} The original Mamba-2 architecture promotes stability by imposing negative scaling on the diagonal of $A$. However, true stability requires all eigenvalues to have negative real parts~\cite{ContwithLLM11}. To guarantee this, stable Mamba-2 modifies each diagonal element of $A$ as:
\begin{equation*}
a_i =
\begin{cases}
a_i, & a_i < 0, \\
-1\times 10^{-5}, & a_i \ge 0.
\end{cases}
\label{eq:stability_condition}
\end{equation*}
This ensures the system matrix is strictly stable, preventing divergence in recurrent dynamics. Replacing zero and positive diagonal entries with a small negative value maintains well-posed dynamics and avoids singularities, resulting in more robust and bounded state trajectories. In general, this family of sparse Mamba models reinforces controllability and observability in the original Mamba model, while enhancing stability in the Mamba-2 model. The experimental results in \cite{ContwithLLM11} show that enforcing controllability, observability, and stability in the Mamba architecture improves hidden-state manipulation, ensures states are fully reflected in outputs, prevents unstable dynamics over long sequences, enhances perplexity and training efficiency, and enables sparse, parameter-efficient models with better generalization and extrapolation. Although SSM-based LLM architectures naturally connect sequential modeling to control theory, their input-dependent dynamics remain theoretically underexplored; however, linear-system insights offer opportunities for explainability and improved design, as demonstrated by models like linear recurrent unit (LRU), highlighting the potential for synergies between LLM research and control-theoretic principles~\cite{ContwithLLM13}.

\section{Challenges and Future Steps}
\begin{figure}
\centering
\includegraphics[width=\columnwidth]{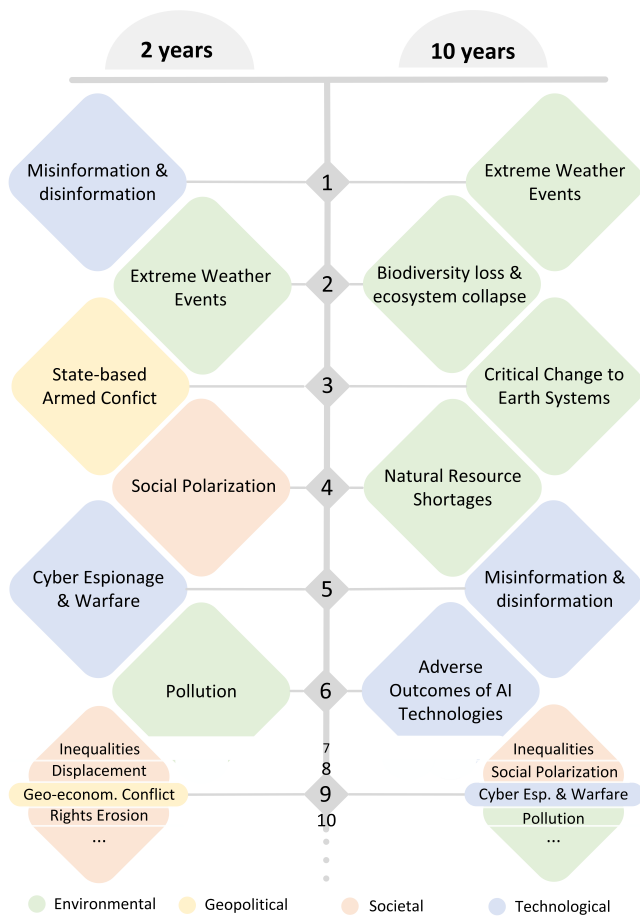}
\caption{The largest risks faced by the world.}
\label{fig_15}
\end{figure}

The rapid proliferation of AI presents both transformative opportunities and significant global risks. As AI-based systems increasingly generate and curate information, concerns related to security, such as bias, disinformation, ethics, and privacy, have become central technological challenges~\cite{Challenges0}. According to the World Economic Forum's Global Risks Report~\cite{Challenges1}, misinformation and disinformation remain among the most severe global risks, ranking fifth over a 10-year horizon, while adverse outcomes of AI technologies rank sixth (see Fig.~\ref{fig_15}). In the context of LLMs and control systems, this motivates a shift from performance-driven adoption toward principled assessment and regulation of LLM behavior. While LLMs can effectively support the design and analysis of control systems, the risks associated with AI-generated misinformation necessitate caution when deploying them, particularly in safety-critical and sensitive applications. Conversely, these risks reinforce the need to apply control-theoretic principles to LLMs to prevent the generation of outputs that may be misleading or misinterpreted as misinformation. Despite early progress, this bidirectional integration of LLMs and control remains in its infancy, motivating the discussion of key challenges and future research trends for developing trustworthy LLM-based systems.

\subsection{Challenges}

\subsubsection{LLM for Control (Indirect)}
LLMs offer substantial potential to enhance control research workflows by supporting literature synthesis, code scaffolding, data preprocessing, and structured report generation. However, current uses are mostly conceptual and illustrative, with little empirical evidence of their impact on research quality. Several major issues remain:
\begin{enumerate}[\textbullet]
\item Limited Empirical Evidence: LLM-assisted workflows in control research are mostly illustrated through isolated examples, with no systematic studies confirming improvements in comprehension or hypothesis generation.
\item Adoption Barriers: While control engineers typically possess the required programming skills, deploying structured or agentic LLM workflows still introduces nontrivial engineering overhead, including prompt design, API integration, and workflow maintenance.
\item Reliability Risks: Unmonitored autonomous use of LLMs may produce hallucinations, inconsistencies, or overconfident outputs, jeopardizing research validity.
\item Cognitive Trade-offs: Excessive workflow automation may diminish critical thinking, conceptual understanding, and deep engagement, even if the tools function correctly.
\end{enumerate}

\subsubsection{LLM for Control (Direct)}
In contrast to indirect support, direct use of LLMs in control embeds language models directly within the control loop. Although still in the early stages, this approach holds promise for reducing human intervention and managing complex specifications while presenting significant challenges:
\begin{enumerate}[\textbullet]
\item System Diversity: Validation is often restricted to narrow benchmarks, simplified laboratory setups, or domain-specific platforms involving low-order linear or mildly nonlinear systems~\cite{DSEddc3,Challenges2}. Evaluations rarely extend to large-scale, safety-critical, uncertain, or highly nonlinear MIMO plants, leaving scalability and transferability to realistic industrial environments unproven.
\item Restricted Control Design: Most approaches focus on classical or narrowly defined controllers~\cite{DSEfeed1,DSEfeed2,EndtoEnd1}, while advanced paradigms, such as constrained optimal, output-feedback, and adaptive designs under complex architectures, remain largely unexplored. Transient performance and hard constraints are usually assessed post hoc rather than incorporated into the design process.
\item Prompt Sensitivity: Frameworks rely heavily on expert-defined prompts, constraints, cost functions, and system abstractions~\cite{Intro30}, making performance highly sensitive and limiting reproducibility and adaptability~\cite{DSEfeed5,Challenges3}.
\item Reliability: LLMs can produce inconsistent reasoning, numerical errors, and biased representations, resulting in unreliable control outcomes~\cite{DSEddc3,DSEfeed2}. Moreover, activation-level interventions can reduce interpretability, making causal understanding and trust in safety-critical applications more difficult~\cite{Challenges4}.
\item Limited Physical Grounding: LLMs lack strong grounding in system dynamics and physical state perception, complicating the translation of high-level reasoning into stable, long-horizon control~\cite{Challenges44,Challenges443}. While hybrid approaches that integrate classical controllers can enhance local decision-making, they do not yet guarantee robust long-term autonomous operation~\cite{Intro19,Challenges444}. Temporal inconsistencies and limited physical awareness remain critical challenges.
\item Robustness: Real-world deployment faces challenges from uncertainties~\cite{EndtoEnd1,Challenges6}. Dependence on commercial APIs adds latency, network failures, and data sovereignty risks. Token context limits constrain multi-turn reasoning and long-horizon planning, and integration with supervisory control systems remains unexplored.
\item Evaluation and Validation: Experiments are mostly limited to simulations or small-scale hardware, with minimal industrial testing~\cite{Challenges5,DSEfeed1,DSEfeed2,Challenges4}. Safety and robustness guarantees remain partial, evaluations often rely on manual inspection, and heavy computational demands and reliance on large foundation models hinder reproducibility and real-time deployment~\cite{Intro30}.
\end{enumerate}

\subsubsection{Control for LLM}
A central limitation in current LLM steering methods is \textit{control strength calibration}, determining when and how much to intervene in model activations. Existing approaches often use fixed intervention strengths, leading to understeering (insufficient error correction) or oversteering (performance degradation). Current methods rely on costly hyperparameter sweeps or ad hoc tuning, and conditional steering techniques are not specifically adapted for diverse error mitigation. Unlike alignment for specific concepts like toxicity, errors are varied and not tied to a single direction in activation space. This raises a fundamental question: \textit{When and how much should models be steered to effectively mitigate errors?} Additional challenges include:
\begin{enumerate}[\textbullet]
\item Limited Safety Guarantees: Open-loop prompting offers no stability guarantees~\cite{Intro27}, and closed-loop activation controllers (e.g., LiSeCo~\cite{Intro2333}) rely on local linear approximations, leaving long-horizon safety unresolved.
\item Applicability: Controllability results are mostly theoretical and overlook emergent behaviors, such as CoT reasoning, creating a gap between theory and practice~\cite{r2777777}.
\item Scalability: Dynamic representation editing~\cite{CtoLLM21} and feedback-based designs~\cite{Intro27} introduce computational overhead, and global guarantees remain unsolved.
\item Detecting Misalignment: Training-time alignment methods (e.g., RLHF) cannot guarantee safety at inference, especially under adversarial prompts. Control-theoretic runtime guardrails~\cite{CtoLLM21,Challenges7} are largely reactive, intervening after unsafe outputs occur rather than preemptively preventing misalignment.
\item Evaluation: Current metrics may not reflect real-world alignment trade-offs, particularly for imbalanced or multi-objective tasks (e.g., toxicity, refusal behavior).
\end{enumerate}

\subsubsection{Control with LLM}
A control-theoretic perspective offers a principled framework for analyzing LLM dynamics via SSMs and Mamba-inspired designs. However, these designs often exhibit recurrent, adaptive, or time-varying behaviors, which complicate the analysis. The key limitations are:
\begin{enumerate}[\textbullet]
\item Complex Latent Dynamics: SSM-based LLMs propagate information through recurrent hidden states, complicating token-level influence and controllability compared to attention-based Transformers. Gradient- and perturbation-based attributions are often unstable, and long-range recurrent interactions obscure input effects, limiting reliable steering and confidence in high-stakes applications.
\item Interpretability: SSM-based LLMs lack explicit attention weights, resulting in limited interpretability and the absence of intuitive control interfaces~\cite{Challenges8}. The implicit state propagation makes it difficult to trace decisions, debug failures, or enforce fine-grained constraints during generation, posing challenges for trustworthy deployment in regulated domains.
\item Reliability: Applying Mamba-inspired architectures to physical systems introduces control challenges, as input-driven selection mechanisms can induce bilinear or time-varying dynamics that complicate stability analysis and controller design. Adaptive representations deviate from classical fixed linear models, raising open questions regarding robustness and formal control guarantees in safety-critical settings.
\end{enumerate}

\subsection{Future Directions}
Building on the identified challenges, several promising research directions emerge across three main categories:
\subsubsection{LLM for Control}
Future progress depends on enhancing how researchers interact with, validate, and trust LLM-augmented systems. Specifically, for indirect LLM applications in control, key directions for future work include:
\begin{enumerate}[\textbullet]
\item Empirical Evaluation: Systematic studies are needed to quantify how LLM support affects comprehension and decision quality across varying automation levels.
\item Human-Centered Interface: Developing intuitive interfaces that lower technical barriers while preserving transparency, reproducibility, and user agency is critical.
\item Lightweight Validation: Integrating real-time consistency checks and error-detection pipelines can improve reliability and foster appropriate trust.
\item Automation Guidelines: Establishing best practices that balance efficiency gains with sustained human oversight and critical thinking is an open research question.
\end{enumerate}
In addition to these directions, standardization through shared benchmarks, datasets, and reporting protocols remains critical for preserving reproducibility and enabling meaningful comparisons across LLM-assisted control research. In the direct perspective, research must push toward broader applicability, reduced expert dependence, and real-world robustness:
\begin{enumerate}[\textbullet]
\item Expanding System and Controller Scope: Future frameworks should address uncertain, stochastic, nonlinear, and large-scale systems. They should also incorporate output feedback, adaptive and robust control, and explicit constraint handling to enhance practical relevance.
\item Minimizing Expert-Based Prompting: A major direction is enabling LLMs to infer control objectives, constraints, and procedures directly from specifications and data, validated through a formal consistency check.
\item Advanced Reasoning and Coordination: Moving beyond ad hoc prompting toward structured reasoning, automated prompt optimization, and multi-agent coordination can improve consistency and reduce cost. Lightweight specialized models may enhance deployability.
\item Vision–Language Models (VLMs): These models offer a promising solution to perception limitations in physical environments, which is critical for realizing the human-centric vision of Industry 5.0~\cite{Challenges443}. By enabling multimodal understanding and embodied interaction, VLMs bridge physical sensing and high-level reasoning. However, challenges remain in robustness, interpretability, long-horizon decision-making, and generalization. Accordingly, key future directions include robust multimodal perception, trustworthy VLMs, long-horizon embodied planning, and Sim2Real generalization for scalable, reliable, and human-centric industrial intelligence.
\item Robust LLM-Based Control: Building on systems like ADAPT~\cite{Challenges44}, future work should enhance motion planning for highly dynamic settings, incorporate semantic environment representations, and safely integrate LLM-enabled agents within control hierarchies. Addressing cloud constraints, extending memory for long-horizon reasoning, and leveraging retrieval-augmented generation, hybrid representations, and physics-informed digital twins will be essential for robust operations.
\item Real-World Validation: Greater emphasis is needed on robustness to noise, disturbances, and modeling errors, along with large-scale benchmarking, hardware-in-the-loop testing, and evaluation in safety-critical settings.
\end{enumerate}

\subsubsection{Control for LLM}
\begin{enumerate}[\textbullet]
\item Optimal Steering: Extending frameworks like Mechanistic Error Reduction with Abstention (MERA)~\cite{Future1} to calibrate steering intensity based on predicted error. Future work should explore safe error correction across broader tasks and alignment objectives, investigating the impact of alternative metrics.
\item Feedback Control Design: Improving feedback-based steering controllers (e.g., RE-CONTROL~\cite{Future11}) for scalability and global guarantees.
\item Formal Guarantees: Extending open-loop prompting with structured metrics and enhancing closed-loop controllers for nonlinear embeddings and long-horizon stability.
\item Scalable Deployment: Ensuring computational efficiency and real-time applicability for high-dimensional LLMs.
\item Preemptive Detection of LLM Misalignment: Extending preemptive control frameworks like reachability-based steering~\cite{Future111} to multi-dimensional notions of harm, improving semantic preservation, and evaluating robustness in diverse, high-stakes real-world applications.
\end{enumerate}

\subsubsection{Control with LLM}
\begin{enumerate}[\textbullet]
\item Control-Aware Latent Dynamics: Developing principled controllability measures for SSM-based LLMs can be an important research direction. Extending frameworks such as the Influence Score~\cite{Future2} to a wider range of models, enabling finer-grained influence tracing, and using anomalous influence patterns for steering can offer clearer insight into what these models learn and how linguistic information is processed internally.
\item Interpretable Control: Future work should create interpretability frameworks and human-readable control abstractions for SSM architectures, using attribution methods and controllable decoding to enable selective steering and constraint enforcement in high-stakes applications.
\item Control-Theoretic Enhancements: Future research should integrate powerful control-theoretic techniques into Mamba-inspired representations of LLMs to enable guided interventions, enforce constraints, and ensure reliable control. Extending Koopman-based solutions (e.g., MamKO~\cite{ContwithLLM12}) with stability and controllability guarantees, along with real-time deployment studies, will be key for safe and practical operation.
\end{enumerate}
\begin{figure}
\centering
\includegraphics[width=\columnwidth]{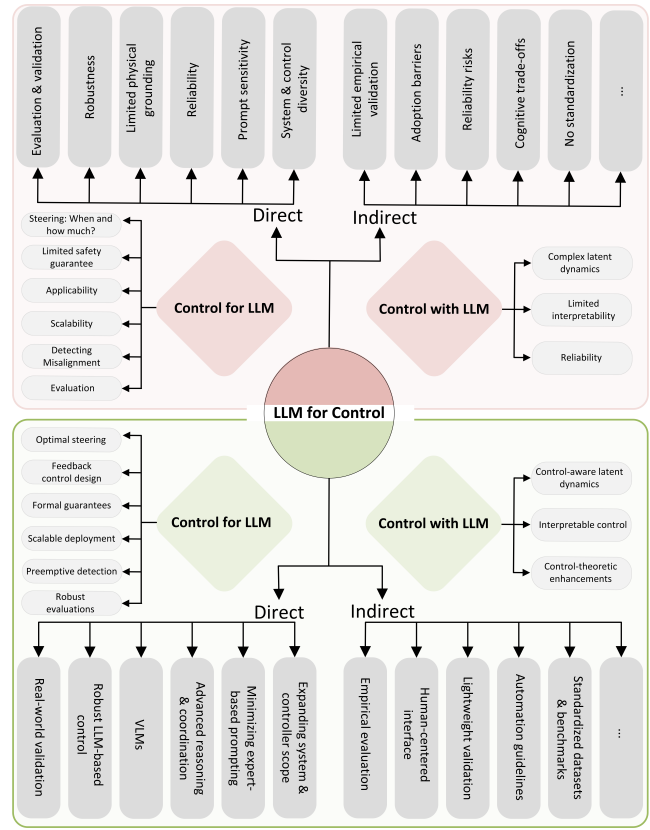}
\caption{Challenges and research opportunities at LLM–control interface.}
\label{fig_16}
\end{figure}

Figure~\ref{fig_16} summarizes the key challenges and future research directions, linking technical limitations to conceptual pathways for LLM-control integration. The highlighted challenges raise a broader question: should LLM alignment be framed solely as a long-term control problem? Classical control theory emphasizes permanent regulation, but this may embed rigid assumptions that scale poorly as model capabilities grow~\cite{Future3}. Trust-based alignment frameworks (e.g., Supertrust) propose an alternative perspective, leveraging temporary scaffolding, cooperative intent, and gradual autonomy. While largely conceptual, these ideas highlight a critical open question for the intersection of control theory and LLMs: ensuring safe and effective AI behavior may require not just stronger control but a reimagining of control strategies themselves.

\section{Conclusion}
In this study, we have examined the emerging interplay between LLMs and control theory, framing it as a bidirectional continuum in which each informs and enhances the other. We highlighted not only how LLMs support control research but also how control principles can steer LLM behavior. By modeling LLMs as dynamical systems, particularly via state-space representations, we emphasized the potential for rigorous internal analysis using concepts such as controllability, observability, and stability. This dual perspective reveals a transformative synergy: LLMs extend the reach and efficiency of control tasks, while control theory provides the tools to regulate, interpret, and guide the behavior of increasingly complex AI systems. Despite these advances, key challenges remain, including model interpretability, the management of emergent behaviors, and the integration of LLMs with physical systems under safety-critical constraints. Addressing these challenges will require interdisciplinary efforts bridging AI, control theory, and dynamical systems.

Ultimately, the convergence of LLMs and control theory promises to produce AI systems that are not only powerful and adaptive but also interpretable, trustworthy, and controllable—qualities essential for their safe deployment in engineering, robotics, and societal applications. Future research at this intersection will be crucial to harnessing the full potential of LLMs while ensuring their alignment with human values and system-level safety.

%\appendix
%\section{My Appendix}
%Appendix sections are coded under \verb+\appendix+.

\printcredits

%\bibliographystyle{cas-model2-names}

\bibliographystyle{IEEEtran}

\bibliography{cas-refs}


%Author biography without author photo.
%Author biography. Author biography. Author biography.
%Author biography. Author biography. Author biography.


%\bio{figs/cas-pic1}
%Author biography without author photo.
%Author biography. Author biography. Author biography.
%Author biography. Author biography. Author biography.

%\bio{figs/cas-pic1}
%Author biography without author photo.
%Author biography. Author biography. Author biography.
%Author biography. Author biography. Author biography.

\end{document}